\documentclass[aps,prd]{revtex4} 
\usepackage[dvips]{graphicx}
\usepackage{amssymb}
\usepackage{amsmath}
\usepackage{natbib}
\newcommand{\be}{\begin{equation}}
\newcommand{\ee}{\end{equation}}
\newcommand{\bea}{\begin{eqnarray}}
\newcommand{\eea}{\end{eqnarray}}
\newcommand{\bitem}{\begin{itemize}}
\newcommand{\eitem}{\end{itemize}}

\newcommand{\benum}{\begin{enumerate}}
\newcommand{\eenum}{\end{enumerate}}
\newcommand{\bc}{\begin{center}}
\newcommand{\ec}{\end{center}}
\begin{document}
\title{WHAT DRIVES OUR ACCELERATING UNIVERSE?}
\author{Sidney Bludman}
\email{bludman@das.uchile.cl} \affiliation{Departamento de
Astronomia, Universidad de Chile, Santiago, Chile}
\date{\today}
\begin{abstract}
The homogeneous expansion history
$H(z)$ of our universe measures only kinematic variables, but cannot
fix the underlying dynamics driving the recent acceleration: cosmographic measurements of the
homogeneous universe, are consistent with either a static
finely-tuned cosmological constant or a dynamic `dark energy' mechanism,
which itself may be material Dark Energy or modified gravity (Dark
Gravity).
Resolving the dynamics of either kind of `dark energy', will
require complementing the homogeneous expansion observations with
observations of the growth of cosmological fluctuations.
Because the 'dark energy' evolution is, at most, quasi-static, any dynamical effects on the fluctuation growth function $g(z)$ will be minimal.  They will be best studied in the weak lensing convergence of light from galaxies at $0<z<5$, from neutral hydrogen at $6<z<20$, and ultimately from the CMB last scattering surface at z=1089.  Galaxy clustering also measures $g(z)$, but requires large corrections for baryonic composition and foreground noise to reduce their large systematic errors.

Projected
observations may potentially distinguish static from dynamic `dark
energy', but distinguishing dynamic Dark Energy from Dark Gravity
will require a weak lensing shear survey more ambitious than any now
projected. Low-curvature modifications of Einstein gravity are also, in principle,
observable in the solar system or in isolated galaxy clusters.

The Cosmological Constant Problem, that quantum material vacuum fluctuations apparently do not gravitate, suggests
identifying gravitational
'vacuum energy' with classical intrinsic spacetime curvature, rather than with any quantum material property. This spacetime curvature of empty space appears cosmologically and about isolated sources and can only be fine-tuned, at present. The Cosmological Coincidence Problem, that we live when the ordinary matter density approximates the 'gravitational vacuum energy', on the other hand, is a material problem, calling for an understanding of the observers' role in cosmology.
\end{abstract}
\maketitle \tableofcontents

\section{INTRODUCTION: COSMOLOGICAL SYMMETRY VS. DYNAMICS}
The most surprising recent cosmological discovery is that the
expansion of our universe began accelerating recently, about redshift $z\sim 0.5$. This paper
will stress the
distinction between the description of the expanding universe
(cosmography or kinematics) and the mechanism driving this accelerating
expansion ('dark energy' or dynamics). Because the homogeneous expansion
$H(z)$ measures only kinematic
variables, it cannot fix the underlying dynamics: cosmographic
measurements of our late accelerating universe, are consistent with
either a {\em static} cosmological constant or a {\em dynamic} `dark
energy', which itself may be a newly-revealed negative pressure material within General Relativity (GR) or modified
gravity (Dark Gravity) \cite{Gu,Starobinsky}.  Indeed, the expansion history probes only one of the gravitational field equations, the Friedmann equation in General Relativity or a modified Friedmann equation in modified gravity.

Starting from the observed global
homogeneity, isotropy and spatial flatness of the universe (flat
Robertson-Walker cosmology, RW), Section II will emphasize the difference between
this RW {\em symmetry} and the {\em dynamics} driving the
cosmological acceleration. Quotation marks are used to stress that `dark
energy' and its 'equation of state' merely summarize
the expansion history. Because we are interested in distinguishing Dark Gravity from Dark Energy, we are careful not to identify RW cosmology with GR, as do many texts and authors.

We want to know whether 'dark energy' is static
or dynamic and whether a newly-revealed material
constituent within General Relativity (Dark Energy), or a
low-curvature modification of General Relativity (Dark Gravity). We will revert to Einstein's original definition
of his cosmological constant $4 \Lambda$ as a {\em classical} intrinsic (Ricci)
spacetime curvature $R_{dS}$ of empty spacetime, the ground state of gravitational theory. Thus, we will interpret the cosmological constant geometrically.
By disconnecting the cosmological constant from material sources, this identification of Dark Gravity with {\em geometry}
avoids addressing the Cosmological Constant Problem, why quantum vacuum energies apparently do
not gravitate in four dimensions.  This Cosmological Constant Problem is clearly a far infra-red gravity problem, not directly connected with four-dimensional quantum gravity. Brane cosmologies invoke extra space dimensions to bridge the huge energy gap between ordinary gravity and other fields, they solve the Cosmological Constant Problem only by fine-tuning the very small scale of these extra dimensions.

Although 'dark energy' may not be
energy at all, but only spacetime curvature, we adhere to common
parlance by denoting as 'gravitational vacuum energy' $\rho_{vac}:=M_P^2 \Lambda=M_P^2
 R_{dS}/4$, the vacuum spacetime curvature $R_{dS}$ or the
cosmological constant $\Lambda$, up to dimensional factors in the Planck mass $M_P$.
This intrinsic spacetime curvature or vacuum energy has generic dynamical consequences which we will discuss in Section III: It distinguishes high- from low-curvature
modifications of Einstein gravity; It modifies the gravitational field surrounding an
isolated source, already at
a {\em Vainstain radius} $r_{*}$, much smaller than the de Sitter
radius $H_{dS}^{-1}$.
These examples will help clarify the physical significance of 'dark energy'.

Section IV will expose the limitations of observations of the
homogeneous expansion: the expansion history observed in the supernova (SN), baryon acoustic oscillation (BAO)) and CMB
is simply
consistent with the Classical Cosmological Constant Model
{$\Lambda$CDM), with a small static cosmological constant $\Lambda$, but
also allows a `dark energy' that is now nearly static. Table III, derived from \cite{Davis,Essence} with some additions in the right column, shows that no present data requires dynamic 'dark energy'. If dynamic, the common 'equation of state' parametrization (11) assumes that it remains quasi-static at high redshifts $z>2$, an assumption that will be tested only after much more gravitational weak lensing (WL) data is observed.

The `dark energy equation of state' and its adiabatic sound speed $c_a^2$ only summarize the homogeneous expansion history $H(z)$, but
cannot resolve the Dark Energy/Dark Gravity degeneracy. However, dynamic
'dark energy' also suppresses the growth of density fluctuations,
which depend on an effective effective sound speed $c_s^2$.  The
difference $c_a^2-c_s^2$ determines the growth function
$g(z):=\delta/a$ of entropic fluctuations
$\delta:=\delta\rho/\rho$. In Section V, we will illustrate how the effective sound
speed and the fluctuation growth function depend on dynamics, by comparing canonical (quintessence) and non-canonical (k-essence) scalar field models of toy Chaplygin gas Dark Energy.

Our formulation stresses the difference between geometry (gravity) and its material sources, and between the Cosmological Constant and the Cosmological Coincidence Problems. If the Cosmological Constant Problem is a problem, it is a problem of empty spacetime curvature: Matter quantum vacuum fluctuations apparently do not gravitate, and the small vacuum spacetime curvature that is observed must be fine-tuned, in any existing theory. On the other hand, the Cosmological Coincidence Problem, why this gravitational vacuum energy and the
material energy densities are just now comparable in magnitude, is a problem for intelligent material observers, which we will discuss in the concluding Section VII.

This formulation stresses the contrived nature of material Dark Energy,
calling it `epicyclic', because new scalar fields are introduced ad hoc, only to
explain dynamically the small present
'gravitational vacuum energy', but still
fail to explain the Cosmic Coincidence.  Low-curvature modifications of classical Einstein gravity, on the other
hand,
are less contrived than fine-tuned Dark Energy, arise naturally
in braneworld cosmology, and may unify early and late inflation. As an alternative to Dark Energy,
we will consider, in Section VI and Appendix C, the Dvali-Gabadadze-Porrati (DGP) braneworld cosmology \cite{DGP,Lue}.

By its nature, the growth function is generally more sensitive to dynamics, than is the homogeneous evolution. Weak lensing
convergence observations of the fluctuation
growth function potentially distinguish static from dynamic
`dark energy' and Dark Energy from Dark Gravity. Indeed,
low-curvature modifications of Einstein gravity, may also be
tested in isolated rich clusters of galaxies
\cite{LueStarkman,Iorio,IorioSecular,Lue}, or even in precision
solar or stellar system tests of anomalous orbital precession or of an increasing
Astronomical Unit. We introduce the Chaplygin Gas Dark Energy and the DGP Dark Gravity models in order to dramatize the necessity of studying the growth of fluctuations. But, because acceptable models are, at most, quasi-static, detecting any small
dynamical effects in the fluctuation growth factor will be difficult: The next decade may distinguish
static from dynamic `dark energy', but will still not distinguish
material Dark Energy from Dark Gravity \cite{Ishak}.

The appendices will summarize present laboratory and solar system tests of General Relativity and
classify Dark Gravity alternatives.

\section{ROBERTSON-WALKER COSMOLOGIES DESCRIBE HOMOGENEOUS EXPANSION }
\subsection{Kinematics: Homogeneity and Isotropy Signify Conformal Flatness} 
Our universe is apparently spatially homogeneous and isotropic, in the large. Such Robertson-Walker cosmologies are
described by the metric \be d s^2=-dt^2+a^2(t)[d
r^2+r^2(d\theta^2+\sin^2\theta d\phi^2)],\\
\ee in which the evolution of the cosmological scale $a(t)=1/(1+z)$ with
cosmic time $t$, is determined by gravitational field
equations, which may be Einstein's or modifications thereof. The comoving volume element is $d a^3$ and other
kinematic observables are listed in Table I, in which overhead dots denote derivatives with
respect to cosmic time.

The Robertson-Walker metric is conformally flat, meaning that it can be rewritten
\be d s^2=a^2[-d\eta^2+d
r^2+r^2(d\theta^2+\sin^2\theta d\phi^2)],\ee
in terms of a conformal or comoving time defined by $a(t)d\eta:= dt$.
This conformal flatness makes light propagate, in every comoving frame, as in Minkowski space, so that the kinematic (geometric) quantities in Table I are observables. Assuming the equation of state $w\geq -1$,
the Weak Energy
Condition $\rho+P\geq 0$ on matter excluding phantom energy in GR, there is no cosmological Big Rip and
the inflationary de Sitter universe $a(t)\sim \exp{H_{dS} t}$ is an attractor for expanding RW universes. The Hubble time then grows continuously:
$\epsilon_H:= dH^{-1}/dt=d\mathcal{H}^{-1}/d\eta+1>0$. But, if the comoving Hubble
expansion rate reaches a minimum, $w$ falls below $-1/3$, the deceleration $q:=(1+3w)/2=d(1/\mathcal{H})/d\eta$
becomes acceleration, and the comoving Hubble
expansion rate starts increasing with comoving time
$d\mathcal{H}/d\eta>0$. This change from deceleration to
acceleration (inflation) happened in the very early universe and again
recently at $z\sim 1/2$ (Figure 1). While early inflation proceeded until the parameter $\epsilon_H\ll 1$, the current
inflation started only recently, so that, although $\epsilon_H <
1$, it does not yet deserve the appellation "slow-roll parameter".

\begin{table}[!b]   
\caption{Kinematic observables for any RW geometry, in terms of
Hubble expansion rate $H:= d\ln{a}/dt$ and comoving expansion rate $\mathcal{H}:=d a/dt$.}
\begin{ruledtabular}
\begin{tabular}[t]{|l|c|}
Description                           &Definition            \\
\hline \hline
Hubble time                         &$H^{-1}:=dt/dN, \qquad N:=\ln{a}$ \\
comoving Hubble time                &$\mathcal{H}^{-1}:= 1/aH =d\eta/dN$ \\
expansion rate of Hubble time       &$\epsilon_H:=dH^{-1}/dt=-d\ln{H}/dN=1-d\ln{\mathcal{H}}/dN=1+q$ \\
cosmological stiffness              &$\gamma(z):=1+w(z):=-d\ln{H^2}/3 dN:=-2\dot{H}/3 H^2:=2\epsilon_H/3$ \\
space expansion                     &$da^3/a^3=3dN=-d\ln{(1+z)^3}=3Hdt=3\mathcal{H} d\eta$\\
\hline
comoving time since Big Bang        &$\eta(z):=\int_0^t dt'/a(t')=\int_z^{\infty}dz'/H(z')$ \\
proper motion distance back to redshift z &$d_M(z)=c\int_0^z dz'/H(z')=c(\eta_0-\eta(z))$\\
deceleration                        &$\ddot{a}/a=H^2+\dot{H},\qquad q:=-\ddot{a}/aH^2=dH^{-1}/dt-1=
d(1/\mathcal{H})/d\eta$ \\
cosmological jerk                   &$\dddot{a}/a=H^3+3H\dot{H}+\ddot{H}, \qquad j:=\dddot{a}/aH^3=1+3\dot{H}/H^2+\ddot{H}/H^3$\\
\hline
spacetime (Ricci) curvature         &$R:= 6(k/a^2+\dot{H}+2 H^2)=6(k/a^2+(\dot{a}/a)^2+\ddot{a}/a)=6(k/a^2+H^2(1-q))$
\end{tabular}
\end{ruledtabular}
\end{table}

\subsection{Dynamics: General Relativity or Modified Gravity } 
In any metric theory, the spacetime curvature (Riemann tensor) would be determined by the matter stress-energy distribution.
But, in a Robertson-Walker universe, conformal flatness implies that all of the Riemann curvature tensor
depends only on derivatives of the Ricci tensor $R_{\mu\nu}$ and ultimately only on the Ricci spacetime curvature scalar. The only gravitational degrees of freedom are those directly connected to matter through the Ricci tensor, which is subject to four differential Bianchi identities $\nabla^\mu R_{\mu\nu}\equiv \nabla_\nu R$, following from general covariance and reducing to six the number of degrees of freedom in $R_{\mu\nu}$ and $g_{\mu\nu}$. Besides the metric $g_{\mu\nu}$, the Einstein tensor $G_{\mu\nu}:=R_{\mu\nu}-g_{\mu\nu} R/2$ is the only covariantly conserved second rank tensor  This does suggests, but does not require, making $G_{\mu\nu}+\Lambda g_{\mu\nu}$ proportional to the matter tensor $T_{\mu\nu}$.

Now, the Hubble expansion rate $H(t)$ derives from the gravitational field equations:  
\begin{description}
\item {\em In Einstein's original General Relativity,} the field equations were   \be G_{\mu\nu}=\varkappa^2 T_{\mu\nu}, \qquad
\varkappa^2:=8\pi G_N :=M_P ^{-2} ,\ee
where $G_N$ is Newton's
constant, and $M_P$ is the reduced Planck mass.  In RW cosmology, there is only one independent field equation, leading to the Friedmann-Robertson-Walker equation
\be H(a)^2:=\dot{a}^2/a^2=\varkappa^2\rho_m/3-k/a^2, \ee
where $\rho_m$ is the material energy density, which
vanishes in empty space.  The vacuum spacetime curvature $R_{dS}=0$.
\item {\em In Einstein-Lema$\hat{i}$itre General Relativity,} the field equations become \be G_{\mu\nu}+\Lambda g_{\mu\nu}=\varkappa^2 T_{\mu\nu} \ee
and the Friedmann-Robertson-Walker equation, the de Sitter radius, and the vacuum spacetime curvature become
\be  H(a)^2:=\dot{a}^2/a^2=(\varkappa^2\rho_m +\Lambda)/3-k/a^2, \qquad H_{dS}^{-1}=\sqrt{3/\Lambda},\qquad R_{dS}=4\Lambda.\ee
For both the original Einstein and the later Einstein-Lema$\hat{i}$itre field equations, the linearity in $G_{\mu\nu}$ guarantees local conservation of matter stress-energy $T_{\mu\nu}$, and the Friedmann-Robertson-Walker equation as their entire content. $H(t)$ is
the only degree of freedom, only the tensor components of the
metric $g_{\mu\nu}$ are propagating, and the Weak Equivalence Principle is satisfied.
\item {\em In Alternative Gravitational Theories,} $T_{\mu\nu}$ is no longer proportional to the
Einstein tensor, the Bianchi identities no longer imply local conservation of matter stress-energy and the modified Friedmann equation now incorporates only one of the independent field equations. $\dot{H}$ and
$\ddot{H}$ become additional degrees of freedom, coupling non-minimally to matter, so that the Weak Equivalence Principle and Newtonian inverse square gravity may no longer be satisfied at short distances.
\end{description}
$H^2$ is linear in the matter energy density $\rho_m$ in GR, non-linear in modified GR.

The important difference between spacetime and spatial curvature $k$ is
illustrated in the two empty stationary RW cosmologies: The flat {\em de Sitter model} has $k=0$, but a
constant spacetime curvature $R=4 \Lambda$ and expansion
rate $H_{dS}=\sqrt{\Lambda/3}$. The {\em Milne
model} $H(a)^2=1/a^2$ has negative spatial curvature $k=-1$, but vanishing
spacetime curvature. The scale is expanding uniformly
$\mathcal{H}=\dot{a}=1$, so that Hubble's original linear relationship between redshift and distance remains exact at all redshifts.

The RW symmetry is broken at small cosmological scales, where
inhomogeneities appear. These inhomogeneities or fluctuations break
translational invariance, leading to Goldstone mode sound waves and
the growth of structure (Section V). This
symmetry-breaking at low temperatures and small cosmological scales
is reminiscent of symmetry-breaking at low energies in condensed
matter and particle physics, except that cosmological structures are gravitationally unstable and will collapse or decay away in an expanding universe.

\subsection{`Dark energy' and `Equation of State' Only Describe the Homogeneous Expansion} 
The expansion history does not determine the dynamics. Although flat Robertson-Walker kinematics does not assume General
Relativity, by defining \be \rho_{DE}:=3 M_P^2H^2-\rho_m, \qquad
w_{DE}(z):=-d\ln{(\rho_{DE}/(1+z)^3)}/3 dN, \ee
the homogeneous expansion history may be parameterized by a
two-component perfect fluid:

\begin{description}
\item [composite mass density:] $\rho:= 3 M_P^2H^2:=\rho_m+\rho_{DE}$
\item [composite pressure:] $P:= -M_P^2 c^2 (3H^2+2\dot{H}):= P_m +P_{DE}$
\item [composite enthalpy density:] $\rho+P/c^2:=-2 M_P^2 \dot{H}=-d\rho/3 dN$
\item [composite fluid stiffness:] $\gamma (z):=-d\ln{\rho}/3 dN:=1+w=-2\dot{H}/3 H^2=\gamma_m\Omega_m+\gamma_{DE}(1-\Omega_m)$
\item [composite `equation of state':] $w(z):=-(3 H^2+2\dot{H})/3 H^2=d\ln{(H^2 /(1+z)^3)}/3dN=w_m\Omega_m+w_{DE}(1-\Omega_m)$
\item [adiabatic sound speed:] $c_a^2:=dP/d\rho:=-(1+\ddot{H}/3H\dot{H})$.
\end{description}
Integrating
$\gamma_{DE}(z)$,
 $\rho_{DE}(z)=\rho_{DE0}(1+z)^{3\overline{\gamma_{DE}(z)})}$,
where $ \overline{\gamma_{DE}}(z):=(1/N)\int_{0}^{N} \gamma_{DE}(N')\,dN'$ is
the past-averaged value of the 'dark energy stiffness', so that
the observed Hubble expansion
 \be
(H/H_0)^2=\Omega_{m0}(1+z)^3 +
(1-\Omega_{m0})(1+z)^{3\overline{\gamma_{DE}(z)})}.\ee  The departure from homologous expansion, or the curvature in the Hubble expansion rate $\mathcal{H}$ (Figure 1) signals the appearance of 'dark energy'. The present acceleration requires
 $\gamma_{DE0}<2/3$, so that the `dark energy' is diluting
slower than the matter density.

Apparently, after a high-curvature early inflationary
phase, our universe expanded monotonically through
radiation-dominated and matter-dominated phases, towards a different
low-curvature late inflationary phase. During each of the four barotropic phases in
Table II, the equation of state, adiabatic sound speed, acceleration, and jerk
were constant and the universe expanded homologously towards smaller spacetime curvature.   During phase transitions that mix these
perfect fluids or introducing cosmological scalar fields,
the `equation of state' changes, the composite fluid is imperfect, entropy is generated, and the expansion is no longer simply homologous. Assuming no phantoms intervene to make $w<-1$, the universe will asymptote monotonically towards a future de Sitter phase of small, but finite, spacetime curvature $R_{dS}$.
\begin{table}   
\caption{Five barotropic phases of our flat universe  expanding homologously as $a\sim t^{1/\epsilon_H}$ or $a\sim\exp{Ht}$, with constant equation of state $w=\gamma-1=2\epsilon_H /3-1$.}
\begin{ruledtabular}
\begin{tabular}{|l||c|c|c|c|c|c||c|}
$\gamma$ &w      &$a(t)\sim t^{1/\epsilon_H}$&H(t)$\sim 1/\epsilon_H t$ &q(t)$=(1+3w)/2$&j(t)$=1+9w(1+w)/2$&$R(t)=6 H^2 (1-q)$                      &Model Flat Universe\\
\hline \hline
4/3      &1/3    &$t^{1/2}$                  &1/2t                      &1              &3                 &0                        &radiation-dominated \\
1        &0      &$t^{2/3}$                  &2/3t                      &1/2            &1                 &$3 H^2$                    &matter-dominated (E-dS) \\
2/3      &-1/3   &t                          &1/t                       &0              &0                 &$6 H^2$                    &coasting \\
1/3      &-2/3   &$t^2$                      &2/t                       &-1/2           &0                 &$9 H^2$                  &accelerating \\
\hline
0      &-1     &$\exp{H_{dS}t}$           &$H_{dS}=const$               &-1             &1                 &$12 H_{dS}^2=4\Lambda$  &inflationary (de Sitter) \\
\end{tabular}
\end{ruledtabular}
\end{table}

 The bottom of Figure 1, from Riess et al. \cite{Riess2}, shows the comoving
Hubble expansion rate $\mathcal{H}=0.68(1+z)^q$ for three
hypothetical phases with constant deceleration $q(z)=0.5$ (upper
dashed curve), constant acceleration $q(z)=-0.5$ (lower dashed
curve)and coasting $q(z)=0$ (central dotted line).
The supernova data is fitted by the central dashed curve $q(z)=-0.6+1.2 z$, changing from deceleration to acceleration
around $z=0.46$ when the comoving
expansion rate reached a broad minimum $\mathcal{H}(z)\sim 0.6$.

\begin{figure} [!t] 
\includegraphics[width=0.8\textwidth,angle=90]{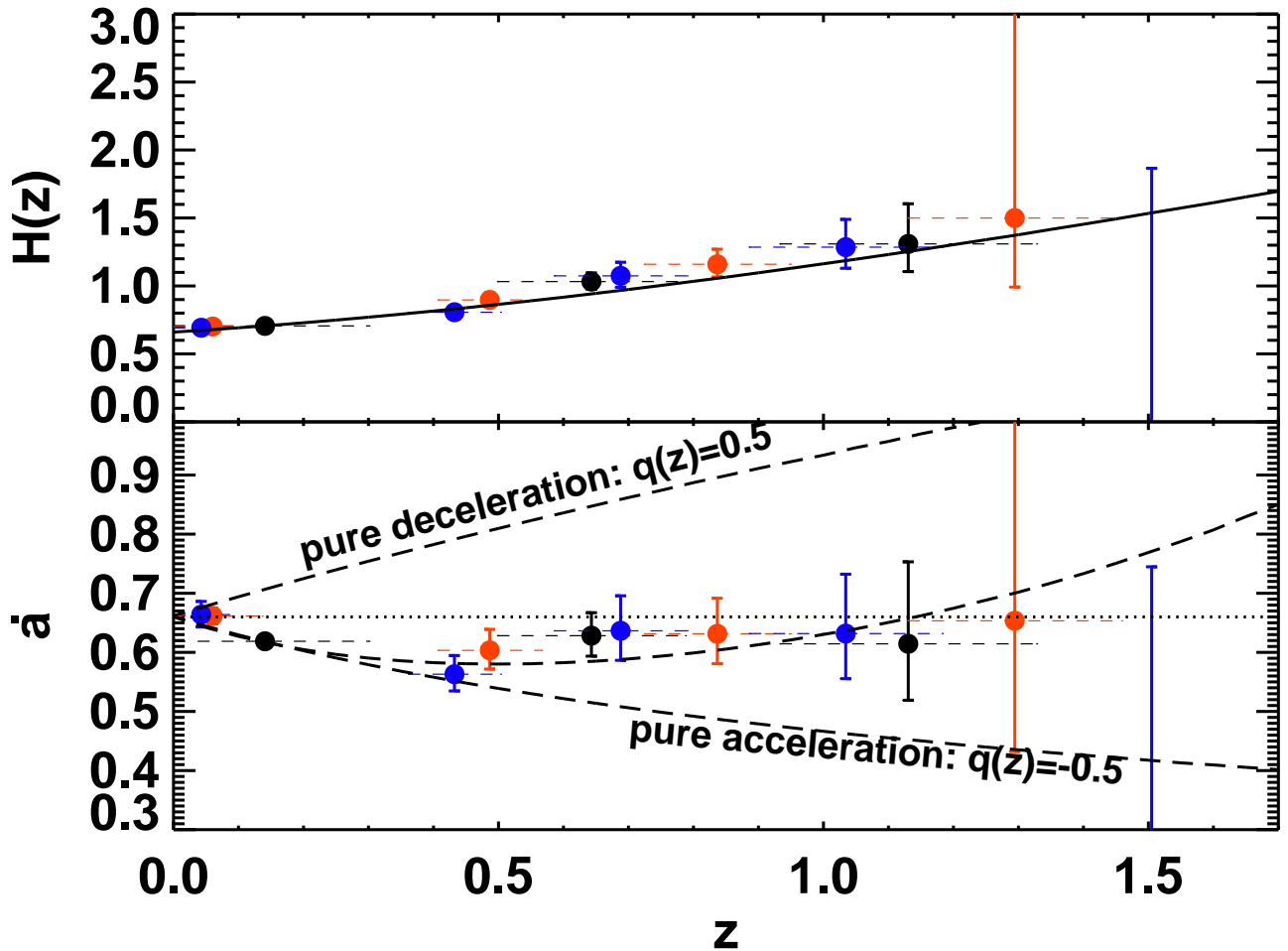}
\caption{Upper panel: Hubble expansion rate, in units of 100
km/sec/Mpc, derived from the supernova Gold sample, compared with
the Concordance Model with
$\Omega_{m0}=0.29,~\Omega_{\Lambda}=0.71$, for which acceleration
starts at $z\sim 0.7$. Lower panel: Conformal Hubble expansion rate
$\mathcal{H}=\dot{a}=H(z)/(1+z)$, compared to constant deceleration
model $q(z)=0.5$ (top dashed curve), coasting model $q(z)=0$ (middle
dotted curve), constant acceleration model $q(z)=-0.5$ (bottom
dashed curve). The Gold data favors the recently accelerating model
$q(z)\approx -0.6+1.2 z$ (middle dashed curve), with acceleration
starting at $z\sim 1/2$ (from Riess et al. \cite{Riess2}, Figure 7).}
\end{figure}
\subsection{'Dark energy' is Now Static or Very Nearly Static} 

By definition,
 $\rho_{DE}(z)$ and $w_{DE}(z)$ simply summarize 'dark energy' and
 its evolution. This 'dark energy' is either
a newly-revealed material constituent within GR or a modification to
GR:

\begin{description}
\item [Dark Energy:] In Einstein's original General Relativity, the total matter stress-energy is covariantly conserved and Friedmann equation (4) is the only independent field equation.
For a given form of
kinetic energy, $\dot{\phi}$ lets the field $\phi (t)$
substitute for the time. If the scalar field is canonical
(quintessence), with kinetic energy density
 $\partial_\mu{\phi}\partial^\mu{\phi}/2:=X/2$, then
 $\dot{\phi}^2=(1+w_{DE})\rho_{DE}$ and the potential energy density
 $V(\phi)=(1-w_{DE})\rho_{DE}/2=(1-w_{DE})(3 M_P^2 H^2-\rho_m)/2$, so that the expansion history
 determines the quintessence potential. If the scalar field is
 non-canonical (tachyonic), the kinetic energy is non-linear in $X$ and $w_{DE}$
 determines a different potential.
\item [Dark Gravity:] If no such Dark Energy exists, then $w_{DE}(z):=w(z)/
(1-\Omega_m (z))$ expresses the Dark Gravity modification to the
Friedmann equation. The simplest form of Dark Gravity is the static Cosmological Constant Model, for which $\rho_{DE}=const,~w_{DE}=-1$. If Dark Gravity is dynamic, then the modified Friedmann equation is only one of the field equations
\end{description}
Dynamical Dark Gravity or Dark Energy introduce scalar fields non-minimally or minimally coupled to gravity. The new scalar fields appear as negative pressure matter in Dark Energy, or as new gravitational scalar degrees of freedom in Dark Gravity i.e. the Friedmann equation (4) is modified on the left side or on the right side.

According to WMAP3 \cite{Spergel}, the present Hubble expansion rate
$H_0=\mathcal{H}_0=73\pm 3~ km/sec/Mpc$, Hubble time
$H_0^{-1}=\mathcal{H}_0 ^{-1}=13.4\pm 0.5~
 Gyr=4.11\pm 0.17~Gpc$, so that the cosmic acceleration has only
increased to $-q_0\approx 0.52$, the "slow-roll" parameter has only
decreased to $\epsilon_{H0}=0.48$, the overall `equation of
state' is now $w_0 \approx -0.74$. For spatially flat $\Lambda CDM$, the SNLS supernova data then implies
$\Omega_{m0}=0.234 \pm 0.035$ and the `dark energy equation
of state' $w_{DE0}=-0.926^{+0.051}_{-0.075}$ \cite{Spergel}.  This
limit to how dynamical the 'dark energy equation of state' can now be
will improve still more, when more weak lensing
measurements of galaxy halo masses and cluster abundances lead to an improved constraint on the matter spectrum
amplitude $\sigma_8$ \cite{clusters}.

The lower curve in Figure 2, from  Koyama and Maartens \cite{Koyama}, shows the Dark Gravity/Dark Energy degeneracy in DGP comoving distance
$d_M(z)$ derived from the DGP expansion history. In Section V, we will see how this Dark Energy/Dark Matter
degeneracy in the homogeneous evolution may be resolved by
prospective observations of the growth of inhomogeneities.
\begin{figure}[!t]
  \begin{center} 
   \includegraphics[width=0.9\textwidth]{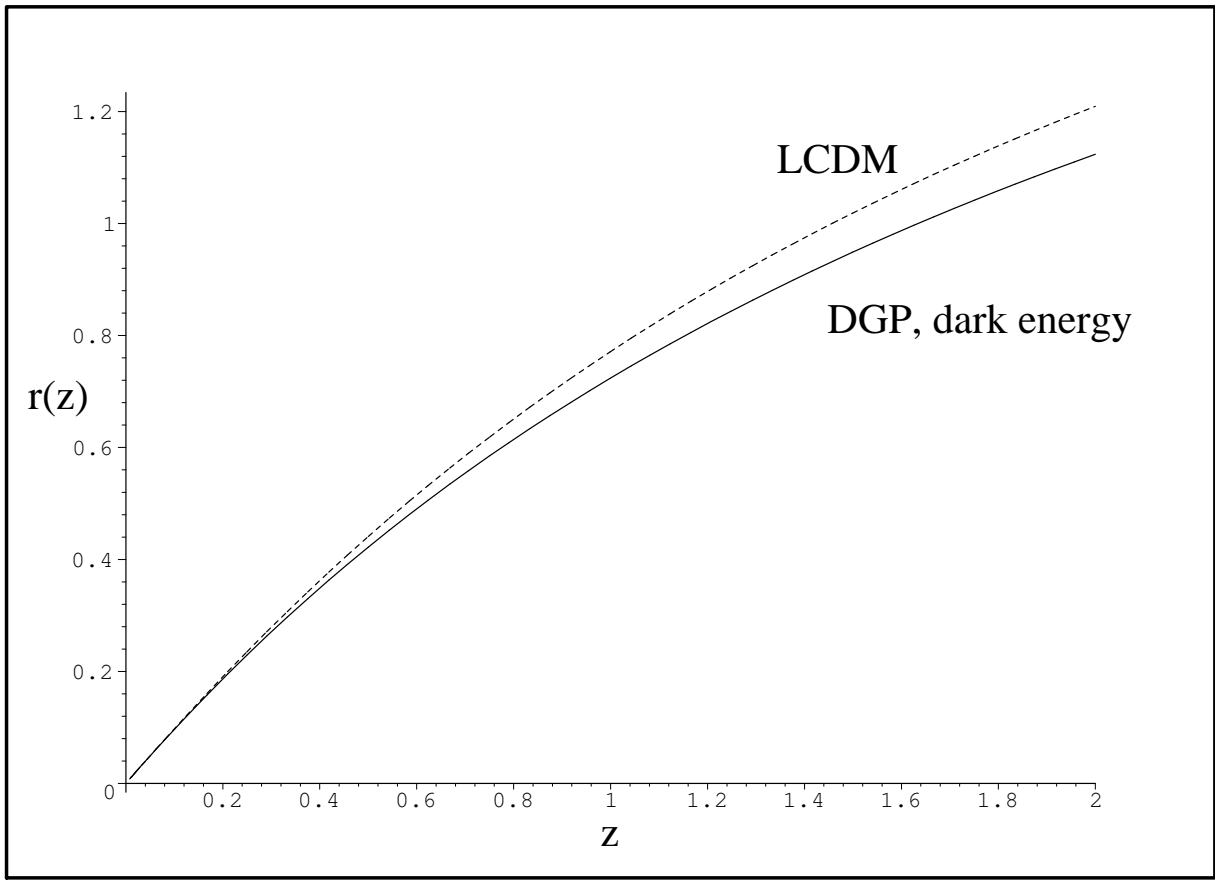}
  \caption{The comoving or proper motion distance $r(z)\equiv d_M(z):=c\int_0^z
    dz'/H(z')$ back to redshift $z$, shows curvature at high redshift depending on the cosmological model.
    Distances obtained from the dynamic DGP Dark Gravity model
    or its equivalent Dark Energy mimic
    are somewhat less than those obtained from
    the static LCDM model because DGP gravity was somewhat stronger in the past.
    (from Koyama and Maartens\cite{Koyama}).}
  \end{center}
\end{figure}

\section{SOME DYNAMICAL CONSEQUENCES OF VACUUM SPACETIME CURVATURE} 
\subsection{Interpretation of Asymptotic Spacetime Curvature As Gravitational Vacuum Energy}
We have been at pains to distinguish between kinematics, as observed in the accelerating expansion of our universe, and the dynamics driving this expansion. Nevertheless, Robertson-Walker symmetry has some dynamical consequences, which help elucidate the physical consequences of vacuum spacetime curvature ('gravitational vacuum energy') and the connection between geometry and material sources. .

In General Relativity, the Bianchi identities assure covariant conservation of the material stress-energy.  This will no longer generally be true for Dark Gravity, where the material stress-energy tensor $T_{\mu\nu}$ is no longer simply proportional to $G_{\mu\nu}+\Lambda g_{\mu\nu}$ and not generally locally conserved. Nevertheless, in empty space, the Ricci scalar and the Hubble expansion rate will still asymptote to the small constant values $R_{dS}=4\Lambda$.
We will interpret Dark Gravity geometrically,
avoiding identifying the vacuum Ricci tensor with any material stress-energy, and
avoiding the Cosmological Constant Problem, by tuning the geometry to $\Lambda:=3 H_{dS}^2=2.19 H_0^2$ or de
Sitter radius $H_{dS}^{-1} \sim 5.2 ~Gpc$. Indeed, there is no evidence that quantum vacuum
energies gravitate at all: the observed electromagnetic
Casimir effect, demonstrates how electromagnetic fields interact with
electromagnetic vacuum quantum {\em fluctuations}, and may have little to
do with the gravitational properties of the vacuum {\em energy}
\cite{Lamoreaux}. The Cosmological Constant Problem expresses the inequivalence of the gravitational and material vacua!

The physical significance this 'gravitational vacuum
energy' will now be illustrated by two examples of how vacuum spacetime curvature
impacts dynamics.
\subsection{Vacuum Curvature Classification of Robertson-Walker Cosmologies}
In the absence of Dark Energy, our accelerating universe is now
dominated by pressure-free matter and Dark Gravity must modify
the Friedmann equation, in
the infra-red. Such low-curvature
modifications preserve Einstein gravity and the Equivalence Principle locally.
Alternatively,
Einstein gravity might be modified in the ultra-violet. Such high-curvature modifications
 require sub-millimeter
corrections to Newton's inverse-square gravity.
In this
way, the presence or absence of vacuum energy distinguishes low- from
high-curvature modifications of General Relativity. While the gravitational vacuum may be intrinsically classical in origin, the high-curvature modifications must involve quantum gravity \cite{Arkani-Hamed,Randall,Binutray}.

\subsection{Vacuum About An Isolated Spherically Symmetric
Source}
The Riemann curvature tensor can always be decomposed into a traceless part (Weyl or conformal tensor) plus a remaining part (Ricci tensor).  In Robertson-Walker cosmologies, the (four-dimensional) Weyl tensor vanishes, the Ricci tensor is determined by the local matter distribution, and the Ricci scalar depends on the acceleration imparted to matter by the field equations. We now consider the opposite extreme, where the the Ricci tensor vanishes or is constant, but the Weyl tensor is determined by the non-local matter sources.

The vacuum energy determines the
gravitational field about any isolated spherically symmetric source, without reference to the source other than its mass $M$ or
Schwarzschild radius $r_S:=2G_N M$:
\bitem
\item In General Relativity, the vacuum field equations are the vanishing
of the Einstein tensor $G_{\mu\nu}:=R_{\mu\nu}-R g_{\mu\nu}/2=0$, and the
unique spherically symmetric vacuum metric is the
Schwarzschild-de Sitter metric \be g_{tt}=g_{rr}^{-1}=1-r_S/r
+\Lambda r^2/3. \ee A vanishing vacuum energy signifies the
Schwarzschild metric: $g_{tt}=g_{rr}^{-1}=1-r_S/r$. This is Birkhoff's Theorem, a generalization of Newton's
Iron Sphere Theorem: for any thin spherical shell, the gravitational
potential vanishes inside, and decreases outside as $1/r$. Birkhoff's clearly {\em
geometric} theorem expresses the significance of vanishing vacuum energy in Einstein gravity.
\item In the Dvali-Gabadadze-Porrati Model,
discussed in Appendix C,
the field equations are the vanishing of
$G_{\mu\nu}+E_{\mu\nu}=0$, where $E_{\mu\nu}$ is the projection of
the five-dimensional Weyl tensor onto the four-dimensional brane,
whose dynamical importance will be apparent in Figure 3. The
DGP vacuum metric is \be g_{tt}=1-r_S/2r+\sqrt{r_S r/2 r_c^2},\quad
g_{rr}^{-1}=1+r_S/2 r-\sqrt{r_S r/8 r_c^2},\qquad r\lesssim
r_{\star}:= (r_S r_c^2)^{1/3},\ee where $r_c$ is a
new cosmological scale. Because this metric differs from the Schwarzschild metric even when $r_c \rightarrow \infty$, there are no DGP Iron Sphere or Birkhoff's Theorems. This emphasizes the characteristic
geometric nature of these theorems in General Relativity
and the distinctive role of the five-dimensional Weyl tensor in brane
cosmology.\eitem

In both cases, the
vacuum energy modifies the Schwarzschild metric at distances beyond the {\em Vainstain radius}
$r_{*}:=(r_S H_{dS} ^{-2})^{1/3}$, where the de Sitter radius $H_{dS}^{-1}$ is $H_0^{-1}$ or $r_c=\beta H_0^{-1}$, for $\Lambda$CDM and for $DGP$ respectively. This
geometric mean between $r_S$ and  $r_c^2$ will also be where
fluctuations start growing according to Friedmann-Lema$\hat{i}$tre or
linearized DGP \cite{LueScoccimaro}, instead of according to
Einstein gravity. These Vainstain scale modifications may
potentially be observable in ultra-precise measurements about isolated Sun-like
stars ($r_S \sim 3~km, ~ r_{\star} \sim 280~pc$) or spherical galaxy
clusters ($r_S \sim 100~pc, ~ r_{\star} \sim 28~Mpc$) \cite{Iorio}.
Their higher-order effects may also someday be tested in
ultra-precise Solar System measurements of anomalous precessions of
planetary or lunar orbits \cite{LueStarkman,DGZ,Sereno} or of a secular
increase in the Astronomical Unit \cite{IorioSecular}.

The homogeneity and isotropy of RW cosmologies implies local isotropy about any point. In a homogeneous universe, what is true locally is true everywhere.
Using Birkhoff's Theorem, Milne and McCrae \cite{Milne,McCrae} were
able to derive the Friedmann equation $k/a^2+H^2=
\varkappa^2\rho/3$ for a pressure-free universe, from Newtonian gravity, without assuming
Einstein's field equations: In a dust universe,
Newtonian cosmology would then have implied the Friedmann equation! (Of
course, in Newtonian cosmology, space would always be flat, so that
the spatial curvature $k$ and scale factor $a$ would lack the
geometrical interpretation GR conveys.) Because vacuum energy is now known to exist, Birkhoff's Theorem
and the Milne-McCrae derivation is today only an
historical curiosity.

In summary, without Dark Energy, the accelerating universe requires vacuum spacetime curvature, called gravitational
vacuum energy.
Besides its cosmological implications, gravitational vacuum energy implies deformation of the
Schwarzschild metric about any isolated source at
the Vainstain radius $r_*$, which is significantly smaller than
the de Sitter radius.

\section{HOMOGENEOUS EXPANSION MEASURES ONLY KINEMATIC VARIABLES} 
While realizing that the homogeneous expansion cannot resolve the
Dark Energy/Dark Gravity degeneracy, we finally review static and dynamic fits to the observed
expansion history.
\subsection{Cosmography: Distances to Supernovae, Luminous Red Galaxies, Last
  Scattering Surface} 

For small radial distances and small galaxy recessional
velocities, Hubble's Law $v=c z=H_0 d$ is a
kinematic consequence of RW symmetry, illustrated by the linear region $d=c z /H_0$ in Figure 2.  But, in
curved spacetime, different global distances are defined
only by physical observables: the comoving distance back to a source
defines the proper motion distance $d_M(z):=c(\eta_0-\eta (z))$; the
observed flux $\mathcal{F}:=\mathcal{L}/4 \pi d_L^2$ from standard candles
defines the luminosity distance $d_L(z):=\sqrt{4 \pi
\mathcal{F}/\mathcal{L}}=(1+z) d_M$;  the angular size $\theta_A$ of
a standard ruler $r_s$ defines the angular diameter distance
$d_A(z):=r_s(z)/\theta_A=d_M/(1+z)$. These different cosmological
distances are observed as follows:
\begin{description} 
   \item[CMB:] The angular diameter distance of the first acoustic CMB peak at the last scattering surface
measures the comoving size subtended at angular scale $\theta_A$.
The measured CMB shift parameter $S:=\sqrt{\Omega_{m0}} H_0
d_M(z_r)=1.70\pm 0.03$ then determines the distance to the last scattering surface at $z_{ls}=1089$ and a standard
ruler, the comoving sound horizon $r_s=147.8^{+2.6}_{-2.7} ~Mpc$ \cite{Spergel,WangMukherjee}.
  \item[BAO:] This provides a
  standard ruler for measuring the line-of-sight distances $H(z)
  r_s$ to luminous red galaxies (LRG)
 and their angular size $d_A(z)/r_s$. From the $z_1/z_{ls}$ distance ratio $R_{0.35}=0.0979\pm 0.0036$ and the measured combination $A:= \sqrt{\Omega_{m0}} H_0 [d_M^2
(z_1)/z_1^2 H(z_1)]^{1/3}=0.469\pm 0.017$, Fairbairn and Goobar \cite{Fairbairn} and Eisenstein et al. \cite{Eisenstein} obtain the
proper motion distance $d_M(z_1)$ to luminous red galaxies, typically at redshift
$z_1=0.35$.
  \item[SN:] The luminosity distances of calibrated supernovae Ia are derived directly from their observed fluxes
  \cite{SCP,SNLS,Miknaitis,Riess2}. The quality data
\cite{SNLS,Miknaitis,Riess2} is now dominated by systematic errors
due to nearby velocity structures and dust.
  \item[WL:] weak gravitational lensing of the light from galaxy
  clusters, from the X-ray emission from hot gravitationally confined electrons, and from the upscattering of CMB relict
  photons by these hot electrons (Sunyaev-Zelovich effect) measures
  the proper motion distances of these sources and the fluctuation
  growth factors $g(z)$ at these distances. Weak lensing
  observations are independent of the baryonic composition of the
  lenses and enjoy a statistical potential far greater than BAO or
  SN.
  \item[GC:] galaxy clustering also measures both $d_M(z)^2/H(z)$
  and $g(z)$, but is subject to large systematic errors, deriving
  from their baryonic composition and foreground noise.
\end{description}      

The proper distance
$c(\eta_0-\eta(z)):=d_M(z)=(1+z)d_A(z)=d_L(z)/(1+z)$, must be
differentiated with respect to redshift, to obtain the Hubble
times $H(z)^{-1}=d\eta/dz,~\mathcal{H}^{-1}=d\eta/d\ln{1+z}$. Obtaining the composite 'equation of state'
$w(z)$ and the 'dark energy equation of state' $w_{DE}(z)=w(z)/(1-\Omega_m(z)$, requires a
second differentiation of the observed distances. This requires
smoothing and binning of the data \cite{Tegmark3}, smears out
information on the `equation of state' \cite{Moar}, and justifies no more than two-
parameter models \cite{LindHut,CaldLinder} in
fitting current and currently underway observations.
The usual Chevalier-Polarski-Linder
parameterization \cite{Polarski2001,Linder} of the 'dark energy equation of state' \be w_{DE}(z)=w_0+w_a (1-a),\qquad
\overline{w_{DE}(z)}=w_0+w_a (1-a) \ln{a} \ee assumes that 'dark energy' grows
smoothly, monotonically and mostly at low redshifts, a prior assumption that will be tested only after more observations at higher redshifts become available. More generally, because observations constrain the
directly observable $H(z)$ and the `dark energy' density $\rho_{DE}$
better than its derivative, it might
be better to parameterize the past average $\overline{w_{DE}}(z)$,
rather than $w_{DE}(z)$ \cite{WangFreese}.

\subsection{Classical Cosmological Constant Model $\Lambda$CDM} 
Einstein introduced his cosmological constant $\Lambda g_{\mu\nu}$
on the left (geometric) side of his original field equations (5),
changing them to the Einstein-Lema$\hat{i}$tre form (5)
and changing the original
Einstein lagrangian $R$ into the Einstein-Lema$\hat{i}$tre
lagrangian $R-2\Lambda$. (Equivalently, the original
Einstein lagrangian $R$ can be varied holding
$\sqrt{g}=-1$. In this unimodular gravity approach
\cite{Buch,Unruh}, $\Lambda$ does not appear in the
lagrangian, but as an undetermined c-number (!) Lagrange
multiplier. This approach stresses the classical nature of the cosmological constant.)

The alternative Dark Matter interpretation subtracts $\Lambda g_{\mu\nu}$ from the right side of equation (9) and interprets the cosmological constant as a constant-density fluid with $\rho_{vac}:=M_P ^2
\Lambda=(2.39~meV)^4$. These two interpretations of
the cosmological constant already exhibit the Dark
Gravity/Dark Energy degeneracy in expansion history.
Although we would prefer to stress the geometric nature of $\Lambda$ as intrinsic spacetime curvature, we will conform to established
parlance by calling it 'vacuum energy'.

Allowing for possible space curvature, the Friedmann equation (6)
contains two
parameters, $\Lambda$ and the present energy density $\rho_{m0}$, which is now almost completely that of non-relativistic matter (dust).  In units of the present critical density $\rho_{cr0}:=3 M_P^2 H_0^2$,
\be (H/H_0)^2=\Omega_{m0}(1+z)^3+\Omega_{\Lambda
  0}+\Omega_{K0},\qquad \Omega_{m0}+\Omega_{\Lambda 0}+\Omega_{K0}\equiv 1,\ee
where $\Omega_{m0}, \Omega_{\Lambda 0}$, are the present matter and vacuum
fractions.
Measuring only the homogeneous expansion cannot resolve this historic static Dark
Gravity/Dark Energy degeneracy. We will now see how this degeneracy
persists, even if the 'dark energy' is made dynamic.

\subsection{Dynamical Cosmological Models}  
The static Cosmological Constant Model can be made dynamic, by
introducing additional parameters relaxing the condition
. Dynamical models allow the vacuum energy to decay down to its present observed
value, no model explains the Cosmic Coincidence, why we are
observing the universe now, when the present matter density
$\rho_{m0}\sim \rho_{vac}/3$. The answer to this question must
involve the observers' role in cosmology (Section VII.B).

Table III, derived from Davis et al.\cite{Davis}, tabulates twelve
`dark energy' fits to the SN+CMB+BAO homogeneous evolution data,
ordered according to the Schwarz Bayes Information Criterion (BIC),
an approximation to the marginal likelihood of improving the fit by
adding more parameters, which measures the strength of each
model in giving the best fit with the fewest parameters.  Although some of these fits derive from interesting Dark Gravity models, their homogeneous evolution can always be mimicked by equivalent Dark Energy models.

The first eight of these twelve models fit the combined data almost equally well.  But,
ordered by BIC, the twelve models fall into four categories of
increasing complexity: \benum
\item The Flat Cosmological Model, the simplest one-parameter fit
to the combined SN+BAO+CMB data, appears on the top row of Table III.  In this model, cosmic acceleration started at $z_{acc}=0.76\pm 0.010, ~6.7\pm 0.4$ Gyr ago, but the the cosmological constant began dominating over ordinary CDM only later at $z_{eq}=0.40\pm 0.04, ~4.3$ Gyr ago \cite{AccelStart}. (All confidence limits are 95\%.). This
`static dark energy` model, with $w_{DE}=-1$,
serves as a standard for comparison with the following eleven dynamical models.
\item The next three models are spatially
curved cosmological constant, spatially flat
constant $w_{DE}$, and flat generalized Chaplygin gas, for which
models, \be P=-A/\rho_{DE}^{\alpha},\quad
(H(z)/H_0)^2=[A+(1-A)(1+z)^{3(1+\alpha)}]^{1/(1+\alpha)},\quad
A^{1/(1+\alpha)}=\Omega_{\Lambda 0}
=1-\Omega_{m0} \ee introduce a
second parameter, without any significant loss in GoF. The broad
uncertainties in the second parameter show the insignificance of
going beyond the simple one-parameter Flat Cosmological Constant
model. In the constant $w_{DE}$ model, cosmic acceleration started at $z_{acc}=0.81\pm 0.30, ~6.8\pm 1.4$ Gyr ago, but the the cosmological constant began dominating over ordinary CDM only at $z_{eq}=0.44\pm 0.20, ~4.5\pm 1.0$ Gyr ago, slightly sooner than in $\Lambda$CDM \cite{AccelStart}.
\item The next four models listed are the variable $w_{DE}(z)$, spatially curved constant
$w_{DE}$, generalized Chaplygin gas \be
P=-A/\rho_{DE}^{\alpha},\quad
(H(z)/H_0)^2=\Omega_{K0}(1+z)^3+(1-\Omega_{K0})[A+(1-A)(1+z)^{3(1+\alpha)}]^{1/(1+\alpha)},
\ee and flat, matter-dominated modified Cardassian polytropic \cite{Freese}, which
expands according to the modified
Friedmann equation\be
(H/H_0)^2=\Omega_{m0}(1+z)^3[1+(\Omega_{m0}^{-q}-1)(1+z)^{3q(n-1)}]^{1/q} .\ee
These four models introduce a third parameter, with insignificant loss in GoF, but at the price of still more complexity.
\item The last four models listed are the spatially flat or curved ordinary
Chaplygin gas and DGP models (17) discussed in Section VI and Appendix C.
The flat models depend on only the one parameter $\Omega_{m0}\approx 0.27$, for which, in the DGP case, $\beta\approx 1.4,~r_c \sim 5.7 ~Gpc$.  The spatially-curved DGP model requires $\Omega_{m0}=0.27\pm 0.03, \Omega_{K0}=0.13\pm 0.05$
These original DGP models cannot simultaneously fit the
SN+BAO and CMB data \cite{Fairbairn} and have poor $GoF\approx 20\%$.
The
two ordinary Chaplygin models simply do not fit all the data.
(Ignoring the BAO and CMB data, Szydlowski et al. \cite{DEfits}
reached the opposite conclusion.) All four DGP and ordinary Chaplygin gas models show too rapid variation of $H(z)$ and are
rejected by their poor GoF. \eenum

\begin{table} [!t] 
\caption{Classical cosmological constant and dynamical
models for the homogeneous evolution. The goodness of fit (GoF)
approximates the probability of finding a worse fit to the data. The
Bayes Information Criteria BIC prefer the one-parameter Flat
Cosmological Constant Model. The $\Delta$BIC values for the other
two- and three-parameter models are then measured with respect to
this Flat Cosmological Constant Model. The table is derived from Davis et al. \cite{Davis}, Table 2, ordered by increasing complexity $\Delta$BIC, but with some additions to the last column.}
\begin{ruledtabular}
 \begin{tabular}{|l||c|c|c|c|}
Model                    &$\chi^2$/dof   &GoF(\%) &$\Delta$BIC &Parameters Fitted\\
 \hline\hline
Flat Cosmologic Constant & 194.5 / 192   & 43.7   &   0        &$\Omega_{m0}=0.27\pm 0.04 $\\
  \hline \hline
Flat Generalized Chaplygin Gas (13)& 193.9 / 191  & 42.7   &   5        &$A=0.73\pm0.05,~\alpha =0.06\pm 0.10 $\\
Cosmological Constant (12)   & 194.3 / 191   & 42.0   &   5        &$\Omega_{m0},\Omega_{\Lambda 0}$\\
Flat constant EOS $w_{DE}$& 194.5 / 191   & 41.7   &   5        &$\Omega_{m0}=0.27\pm 0.04,~w_{DE}=-1.01\pm 0.15 $ \\
\hline
Flat variable $w_{DE}(z)$ (11)& 193.8 / 190   & 41.0   &  10        &$\Omega_{m0}=0.27\pm 0.04,~w_0=-1.0\pm 0.4,~w_a=-0.4\pm 1.8$\\
Spatially-curved constant $w_{DE}$& 193.9 / 190& 40.8&10       &$\Omega_{m0},~w_{DE},~\Omega_{\Lambda}$ \\
Generalized Chaplygin Gas (14)& 193.9 / 190   & 40.7   &  10        &$A, \alpha, \Omega_{K0}$\\
Cardassian Polytropic  (15)   & 194.1 / 190   & 40.4   &  10        &$\Omega_{m0},q,n$ \\
  \hline\hline
Flat Dvali-Gabadadze-Porrati (17)               & 210.1 / 192   & 17.6   &  14        &$\Omega_{m0}\approx 0.27$\\
Dvali-Gabadadze-Porrati (17)  & 207.4 / 191   & 19.8   &  18        &$\Omega_{m0}=0.27\pm 0.05,~\Omega_{K0}=0.13\pm 0.02$\\
Ordinary Chaplygin Gas (14,$\alpha=1$)       & 220.4 / 191   &  7.1   &  30        &$A,~\Omega_{K0}$\\
Flat Ordinary Chaplygin Gas (13,$\alpha=1$)  & 301.0 / 192   &  0.0   &  30        &$A$\\
\end{tabular}  \end{ruledtabular} \end{table}

Acceptable models all agree at low
redshift and are now static or quasi-static. Table III excludes fast evolution, such as would be predicted by many  Dark Gravity models. For example, in Figure 2 \cite{Koyama}, the slopes $d_M/dz=H(z)^{-1}$
show that the too-dynamic ordinary
DGP model would predict expansion rates, that already at $z=2$, evolve about $10\%$ faster than those obtained from the static Classical Cosmological Constant Model.

The simplest and best fit to all the combined is the Flat Cosmological Constant Model on the first line of Table III. The remaining seven acceptable quasi-static models on lines 2-8 of Table III, with essentially the same GoF as $\Lambda CDM$, have their additional parameters so poorly constrained so that, in
the worst cases, Davis et al. \cite{Davis} did not quote their
values. These complex models, with more
parameters, will only be tested after much
more high redshift supernova or weak lenses are observed \cite{Riess2}. Although no evidence requires any dynamical model for
'dark energy', until these seven models are
excluded observationally, we will go on to study possible dynamical
manifestations of gravitational vacuum energy.

\section{GROWTH OF FLUCTUATIONS CAN DISTINGUISH COSMODYNAMICS} 
Dynamical 'dark energy' has two distinct effects: it alters the homogeneous evolution $H(z)$,
as already discussed, and it alters the growth of fluctuations through the
fluctuation {\em growth factor}, which we now discuss.
\subsection{Dark Energy: Canonical and Non-canonical Scalar Fields (Quintessence and
K-essence)} 
Dark Energy and its alternatives are reviewed in \cite{DETF,Padmanabhan}
and ten model fits to the expansion history, with and without
spatial curvature and cosmological constant, are tabulated by
\cite{DEfits}. If Dark Energy exists, it is usually attributed to an
additional ultra-light scalar field $\phi$ with lagrangian $\mathcal{L}(\phi,X)$, where
$X:=\partial_\mu{\phi}\partial^\mu{\phi}$.  The pressure, energy density, `equation of state', and adiabatic sound speed are then
$P=\mathcal{L},~\rho=2 X
\mathcal{L}_{,X}-\mathcal{L},~w=P/(2 X
P_{,X}-P),~c_a^2:=dP/d\rho$ respectively. The lagrangian $\mathcal{L}$ is canonical ({\em quintessence\/}) or non-canonical
 ({\em k-essence\/}) according to whether the kinetic energy
 is linear or not in $X$:
 \begin{description}
 \item [Quintessence] is driven by a slow-rolling scalar potential that can be tuned
to track the radiation/matter until it now dominates, with 'equation of state' now decreasing
$dw_{DE}/dz>0$. In any canonical scalar field, inhomogeneities will propagate with an effective sound speed $c_s^2=1$.
\item [K-essence] is kinetic energy driven, can be chosen to track only the radiation energy density,
so that, after radiation/matter equality, k-essence can start dominating over ordinary matter. The  $w_{DE}$ dropped sharply
near matter/radiation equality and has been increasing thereafter $d
w_{DE}/dz<0$. K-essence apparently cannot arise as a low-energy
effective field theory of a causal quantum field theory
\cite{Durrer}.
\end{description}

Dynamical Dark Energy was originally invoked to
make a material vacuum energy decay down to the present small
value $\rho_{vac0}\sim 2 M_P^2 H_0^2  \lll M_P ^4$ or small expansion rate $H_0^2 \lll M_P^2$.
Both kinds of Dark Energy ultimately require
fine-tuning in different ways: quintessence, in order to make tracing stop before now; k-essence, in order to initiate the transition towards
dominance in the matter-dominated epoch.
Because these scalar fields are non-renormalizable and fundamentally
unnatural, both need to be interpreted as ad hoc low-energy effective
field theories.

\subsection{Entropic Fluctuations Determine the Growth Factor} 
In a mixture of cosmological fluids or with dynamical scalar fields, the
`equation of state' is generally not adiabatic: fluctuations
propagate with an {\em effective
sound speed squared\/} $c_s^2:=P_{,X}/\rho_{,X}=(1+2(P_{,XX}/P_{,X})
X)^{-1}$, which equals unity only in a canonical field theory. The entropic
pressure fluctuations are proportional to $(1+w)(c_s^2-c_a^2)$, so that in the quasi-static limit $w(z) \sim -1$, they are small and
insensitive to the effective sound speed. This minimizes the
differences between static and dynamic Dark Energy
fluctuation growth factors that we would like to distinguish.

How different microscopic dynamics and different effective sound speeds can underly the same Dark Energy
`equation of state' is illustrated by the original Chaplygin gas
\cite{Bento}, whose adiabatic `equation of state' $P(\rho)=-A/\rho$
and adiabatic sound speed $c_a^2=-P/\rho=-w(z)$. This equation of state can be derived from either a
non-canonical or a
canonical scalar field. If
derived from the constant potential Born-Infeld Lagrangian
$\mathcal{L}= -V_0 \sqrt{1+\varkappa^2 X}$, with
non-canonical $\rho=V_0/\sqrt{1-\varkappa^2
X},~P=-V_0\sqrt{1-\varkappa^2 X},~V_0^2\equiv A$, the
fluctuations remain adiabatic, and the effective sound speed
$c_s^2=c_a^2=-w(z)$. But, if derived from the canonical scalar field
with potential $V(\phi)=(\sqrt{A}/2)[\cosh{3\phi}+1/\cosh{3\phi}]$ \cite{Bento}, entropic fluctuations make the effective sound speed $c_s^2=1\neq c_a^2$.
These canonical and non-canonical Chaplygin gas models give the same
adiabatic sound speed $c_a^2=-w(z)$, but different effective sound
speeds, $c_s^2=1,~-w(z)$ respectively, and different fluctuation growth factors. (As mentioned in
Section IV.C, this original Chaplygin gas does not fit
the observed expansion history, but can be generalized to
$P=-A/\rho^{\alpha},~c_a^2=-\alpha w,~\alpha=0.06\pm 0.10$, which is indistinguishable from $\Lambda CDM$ and
fits very well \cite{Amendola2003,Bento,Zhu}.)

\section{MODIFIED GRAVITY: DVALI-GABADADZE-PORRATI BRANE COSMOLOGY} 

Because Dark Energy is contrived, requires fine tuning and
cannot be directly detected in the laboratory or solar system, we
now turn to Dark Gravity as the alternative dynamical source of
cosmological acceleration. These Dark Gravity alternatives,
classified in the Appendix, arise naturally in braneworld theories,
naturally incorporate a small spacetime intrinsic curvature, and may
unify `dark energy' and dark matter, and early and late inflation.
While fitted to the observed cosmological acceleration, they may
also ultimately be tested in the solar system, Galaxy or galaxy
clusters \cite{LueStarkman,Iorio,IorioSecular,Lue,Sereno}.

In the self-accelerating solution of the original DGP model
\cite{DGP,Deffayet2001,Deffayet2}, discussed in Appendix C, gravity leaks into the five-dimensional bulk
at cosmological scales greater than $r_c$,
weakening gravity
on the four-dimensional brane. This leads to a four-dimensional
Friedmann equation \be H^2+k/a^2-H/r_c=\varkappa^2\rho/3 .\ee
Defining $\beta:=r_c H_0$, which can be written as \be
(H(z)/H_0)^2=[1/2\beta+\sqrt{(1/2\beta)^2+\Omega_{m0}
  (1+z)^3}]^2+\Omega_{K0}(1+z)^2,1\equiv\Omega_{m0}+
\Omega_{K0} +\sqrt{1-\Omega_{K0}}/\beta , \ee interpolating between a past matter-dominated universe and a future de
Sitter universe.

Fig. 3 shows the growth factors in this DGP Dark Gravity and its Dark Energy mimic both evolving substantially faster than in the Cosmological Constant Model $\Lambda CDM$. This figure also shows a relatively smaller difference between Dark Gravity DGP and its Dark Energy mimic. In the next decade, weak lensing observations may distinguish between static and dynamic 'dark energy', but not between Dark Gravity and Dark Energy.

As discussed in Section IV.C, the original DGP models cannot simultaneously fit the SN+BAO and the CMB data.
We have used these models only to suggest the importance of
the 5D Weyl tensor in any braneworld dynamics and to illustrate how dynamics is better tested in the growth of fluctuations (Figure 3) than in the homogeneous expansion history (Figure 2). In any realistic model, because the evolution is, at most, quasi-static, any dynamical effects on the growth of fluctuations will be minimal, and will be best studied in the weak lensing convergence of light from galaxies at $0<z<5$, from neutral hydrogen at $6<z<20$, and ultimately from the CMB last scattering surface at z=1089
\cite{DETF,Ishak}.  Galaxy clustering also measures $g(z)$, but requires large corrections for baryonic composition and foreground noise to reduce their large systematic errors.

\begin{figure}[!t]
  \begin{center} 
  \includegraphics[width=0.9\textwidth]{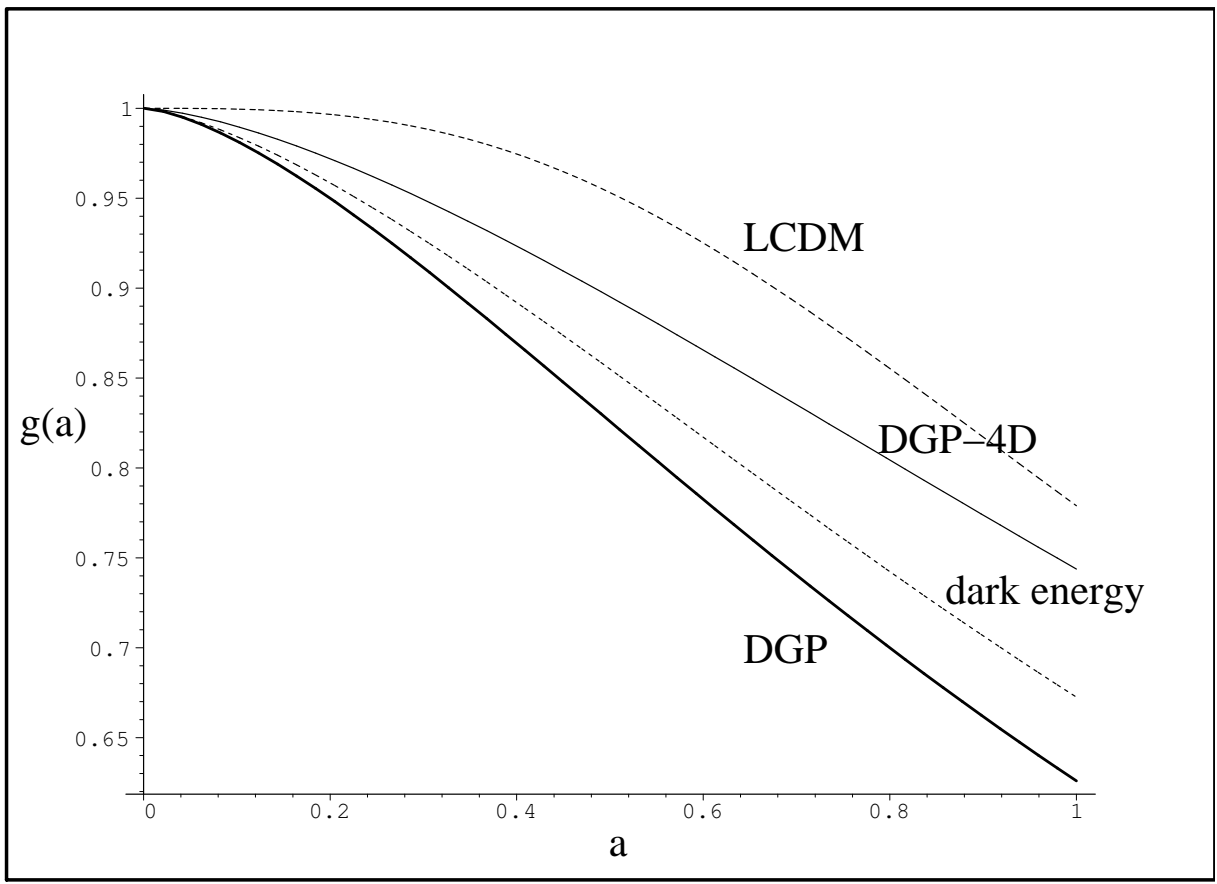}
  \caption{The linear growth history $g(a):=\delta/a$ for flat $\Lambda$CDM
    (long dashed), DGP Dark Gravity (thick solid)
    and Dark Energy (short dashed)
    models in Fig. 2.  DGP-4D (thin solid) shows the incorrect
    result that would be obtained by neglecting perturbations of the DGP
    5D Weyl tensor. The 5D Weyl tensor perturbations thus
    distinguish the static Dark Energy model from the two dynamical Dark Gravity and mimicking Dark Energy
    models. In the DGP Dark Gravity model, Newton's `constant'
    weakens with time, so that its growth suppression evolves even faster than
    in its mimicking Dark Energy model. (from \cite{Koyama} ).
    }
\end{center}
\end{figure}
\section{COSMOLOGICAL CONSTANT, FINE-TUNED DARK ENERGY, OR MODIFIED GRAVITY?} 

\subsection{Phenomenological Conclusions: Vacuum Energy is Now Static or Quasi-static}
We have reviewed present and prospective observations of 'dark
energy', in order to emphasize the differences between kinematical
and dynamical observations, between static and dynamic `dark
energy', and between Dark Energy and Dark Gravity. We conclude:

\bitem
\item Cosmological acceleration is explicable by either a small
  fine-tuned cosmological constant or by `dark energy', which is now
  nearly static.  If dynamic, this `dark energy' is
  either an additional, ultra-light negative pressure material within General
  Relativity, or a low-curvature modification of Einstein's field
  equations.
\item The simplest and best fit to the expansion history, the Classical Cosmological Constant Model,
interprets 'dark energy'
as a classical intrinsic
spacetime curvature, giving geometric structure to empty space.  This
classical interpretation distinguishes 'gravitational vacuum energy' from the ground state of quantum matter, and renounces any attempt to explain its small
value as a quantum vacuum energy.
\item The observed homogeneous expansion history may also be fitted by
  `dark energy' decaying from its huge primordial value and now nearly static. This static or quasi-static 'dark energy' presently observed, whether Dark Energy or Dark Gravity requires
  fine-tuning.
\item  The
  inhomogeneity growth rate potentially distinguishes between static and
  dynamic `dark energy' and between Dark Energy and Dark Gravity. Because the 'dark energy' is now static or nearly static, differences in the
  large-scale angular power spectrum, mass power
  spectrum, or gravitational weak lensing \cite{CFHT} will be small, but may distinguish
  static from dynamic `dark energy'. Distinguishing between Dark Energy and
  Dark Gravity will remain more problematic.
\item No model yet explains the
Cosmological Constant Problem, why quantum vacuum energies
apparently do not gravitate. Nor does any model explain the Cosmic Coincidence, why we observers live at a time when the matter and vacuum energy densities are comparable.
\eitem

Low-curvature modifications of Einstein gravity are conceptually less contrived than fine-tuned Dark
Energy, explain cosmological acceleration as a natural consequence
of geometry, and may unify early and late inflation. Geometric
modifications may be intrinsic in four dimensions, or may arise
naturally in braneworld theories. Invoked in the first
place to explain recent cosmological acceleration, these low-curvature
modifications of Einstein gravity may even be testable by refined
solar system or galaxy observations.  The outstanding problem in both Dark Energy or Dark Gravity remains the significance of the Cosmic Coincidence, which (unless it is fine-tuned), clearly refers to the role of conscious observers.

\subsection{Metaphysical Conclusions: The Role of Observers}

The observed cosmological acceleration requires a small gravitational vacuum energy, which is now static or nearly static, and inequivalent to the material vacuum energy (Cosmological Constant Problem}.
This material vacuum energy is now, when we are observing, comparable to the observed gravitational vacuum energy or cosmological constant (Cosmic Coincidence Problem).

Cosmology is a science, whose observations are
confined to our past light cone and by the size of the universe, and
constrained by cosmic variance \cite{Ellis}.  It differs from other, simply descriptive physical sciences, by its evolutionary character, which may call for a new selection principle among possible universes or different cosmological constants.
The expanding scope of physics has always required new paradigms, such as the Relativity, Equivalence, Complementarity, and Uncertainty Principles,
limiting what can be observed. It should, therefore, come as no surprise if the Cosmic Coincidence calls for a new ontological principle, limiting what can observed about the gravitational vacuum.

Unless a fundamental theory explaining Cosmic Coincidence can be discovered, our own gravitational vacuum is selected by the presence of observers (Weak Anthropic Principle).
This selection may apply only to the cosmological constant and/or only to our own observable universe. It
may select among an ensemble of conceivable theories for our universe \cite{BludRoss} or among an ensemble of real universes, which may exist now as subuniverses of a megauniverse (landscape), may recur periodically
in a bouncing universe, or may evolve from one universe to another by natural selection \cite{Smolin}.

We are only now beginning to understand how observers perceive and interpret reality.
The mind is the window to reality, but our interpretations of reality always depend on past experience, and are constrained by the structure of our brains, in which
cognition, consciousness and feeling are only now becoming physical
observables. Perhaps all aspects of
reality, including observers' cognition, will ultimately be
reducible to known physical principles.  But until then,
the choice between strict reductionism and a new
cosmological principle remains a subjective choice between still-hopeful
string theorists and more skeptical physicists \cite{WeinbergAnthro}.
\appendix
 \section{HOW CAN GENERAL RELATIVITY BE MODIFIED?}
 General Relativity is a rigid metric structure incorporating
general covariance (co-ordinate reparametrization invariance), the
Equivalence Principle, and the local validity of Newtonian gravity
with constant $G_N$, in the weak field and non-relativistic limits.
General covariance implies four
Bianchi identities on the Ricci curvature tensor. The linearity of the Einstein-Hilbert action
in the Ricci scalar curvature, makes the Einstein field equations second order,
the two tensorial (graviton) degrees of freedom dynamic, and
constrains the scalar and vector $g_{\mu\nu}$ degrees to be
non-propagating.

General Relativity differs from Newtonian cosmology only by pressure
or relativistic velocity effects, which are tested in the solar
system and in cosmology (gravitational lensing of light, nucleosynthesis, dynamical age, large
angular scale CMB, late-time mass power spectrum).  Therefore,
modifications of General Relativity must be sought, in order of
scale: in laboratory violations of the Equivalence Principle
(E$\ddot{o}$tvos experiments); in solar system tests
(lunar ranging, deflection of light, anomalous orbital precessions
of the planets and Moon) \cite{Damour,LueStarkman,Lue,Gabadadze,Sereno},
secular increase in the Astronomical Unit \cite{Iorio}); in galaxy
and galaxy cluster number counts \cite{IorioSecular,Sereno2006}; in
gravitational weak lensing; in cosmological variation of Newton's
$G_N$ and other 'constants'; in the enhanced suppression of
fluctuation growth, on large scales or at late times.

Because in General Relativity only the metric's tensor degrees of freedom are propagating, modifying the lagrangian $R$
introduces additional scalar and vector degrees of freedom,
represented by scalar or vector gravitational fields.
The basic distinction between high- and low-curvature modifications
of General Relativity depends on the spacetime curvature of their
vacua. While high-curvature (ultra-violet) modifications have always
been motivated by quantum gravity, low-curvature (infra-red)
modifications are now motivated by the discovery of the recently
accelerating universe and apparently now quantum in origin. Ultra-violet and infra-red modifications both still present fundamental
theoretical problems which we will ignore, since our focus is on
phenomenology.

\section{FOUR-DIMENSIONAL MODIFICATIONS OF GENERAL RELATIVITY} 

For historical and didactic reasons, we begin by summarizing
four-dimensional metrical deformations of General Relativity, which
often appear as projections of higher-dimensional theories, inspired
by string theory \cite{DamourPolyakov,DamourPolyakov2}. \bitem
\item Scalar-tensor gravity, the oldest and simplest extension
  of General Relativity \cite{Fujii,Cap2005}: In the original Jordon
  lagrangian, a scalar gravitational field, proportional to time-varying
  $1/G_N$, couples linearly to the Ricci scalar $R$. After a
  conformal transformation to the Einstein frame, the scalar
  gravitational field is non-minimally
  coupled to matter, so that test particles
  do not move along
  geodesics of the Einstein metric. Instead, test particles move along
  geodesics of the Jordon metric, so that the Weak
  Equivalence Principle holds \cite{Starobinsky2000}.

  Scalar-tensor theories modify Einstein gravity at all scales and
  must be fine-tuned, to satisfy observational constraints.
  Nucleosynthesis and solar system constraints severely restrict any
  scalar field component, so that any Dark Gravity effects on the
  CMB or cosmological evolution must be very small.
  \cite{Bertotti,Cap2006,Catena}.
\item In higher-order $f(R)$ theories, the lagrangian is
no longer simply linear in the Ricci scalar $R$, so that the equations of motion
become fourth-order, equivalent to scalar-tensor theories. f(R) theories is liable to either
negative kinetic energies or negative potential energies. Negative
kinetic energies are unavoidable if the lagrangian depends upon
higher order curvature invariants, such as $P\equiv
R_{\mu\nu} R^{\mu\nu}$ or $Q\equiv
R_{\alpha \beta \gamma \delta} R^{\alpha \beta
\gamma \delta}$ \cite{Ostrogradski,Woodard2006}. The same kinetic
instability afflicts lagrangians involving derivatives of any
curvature scalar, except total derivatives such as the
Gauss-Bonnet invariant, which can
be eliminated by partial integration. If the lagrangian is
restricted to be a nonlinear function of the Ricci scalar, the
resulting kinetic energies are positive.

The simplest low-curvature modification, replacing the Einstein
lagrangian density by $R -\mu^4/R$
\cite{Carroll,Carroll2,Nojiri}, leads to accelerated expansion at
low curvature $R \leq \mu^2\sim H_0^2$. However, outside matter,
this theory is weakly tachyonically unstable and phenomenologically
unacceptable \cite{Soussa}. Inside matter, this tachyonic
instability is vastly and unacceptably amplified \cite{Dolgov}. These
potential instabilities are, however, not generic and one can hope
to avoid them by fine tuning the dependence upon $R$. This
is not surprising, because all $f(R)$ theories are
equivalent to scalar-tensor theories with vanishing Brans-Dicke
parameter $\omega_{BD}=0$ \cite{Teyssandier,Olmo,CapNojOdin},
which can also be fined-tuned to avoid potential instabilities and
to satisfy supernova and solar system constraints
\cite{Odintsov2,Odintsov3,Soussa,Woodard2006}, but not cosmological
constraints \cite{Amendola2}.
\item TeVeS (relativistic MOND theory): Adding an additional vector
  gravitational field, could explain flat galactic rotation curves and the
  Tully-Fisher relation, without invoking dark matter, and could possibly
  unify dark matter and `dark energy' \cite{Bek}. Because gravitons and
  matter have different metric couplings, TeVeS predicts that
  gravitons should travel on geodesics different from photon and
  neutrino geodesics, with hugely different arrival times from
  supernova pulses. It also predicts insufficient power in the third
  CMB acoustic peak \cite{Skordis}. In any case, now that WMAP3 data
  requires dark matter \cite{Spergel}, the motivation for TeVeS
  disappears.
\eitem

\section{EXTRA-DIMENSIONAL (BRANEWORLD) MODIFICATIONS} 

In extra-dimensional braneworld theories, scalar fields appear
naturally as dilatons and modify Einstein gravity at high-curvature,
by brane warping \cite{Randall,Binutray}, or at low-curvature, by
brane leakage of gravity \cite{DGP}.  If quantized, these theories
encounter serious theoretical problems (ghosts, instabilities,
strong coupling problems) and are not now derivable from fundamental
quantum field theories. Until these problems can be overcome, these
theories must be regarded as effective field theories,
incorporating an extremely low infra-red scale at low
spacetime curvature, unlike other effective field theories which
incorporate ultra-violet parameters.

In the original DGP model \cite{DGP,Deffayet2001,Deffayet2}, leakage of gravity into the five-dimensional bulk
leads to a Friedmann
equation, \be H^2+k/a^2-H/r_c=\varkappa^2\rho/3 ,\ee
modified on the four-dimensional brane,
by the additional curvature term $H/r_c$ at the
cosmological scale $r_c$. In our matter-dominated epoch, this modified Friedmann equation is
\be (H(z)/H_0)^2=[1/2\beta+\sqrt{(1/2\beta)^2+\Omega_{m0}
  (1+z)^3}]^2+\Omega_{K0}(1+z)^2,1\equiv\Omega_{m0}+
\Omega_{K0} +\sqrt{1-\Omega_{K0}}/\beta, \ee
where the terms inverse in $\beta:=H_0 r_c$
express the weakening of gravity on the brane at large scales $a>r_c$, due to leakage
into the five-dimensional bulk. This modified
Friedmann equation interpolates between the past matter-dominated universe,
for small
scales $a\ll\beta^{2/3}$, and the future
de Sitter universe with constant
Hubble expansion $H_{dS}:= 1/r_c$, for scales $a\gg 1$.

For the intermediate value $\beta=1.39,~r_c\sim 5.7~Mpc$, the
universe began its late acceleration at
$z_{acc}=(2\Omega_{m0}/\beta^2)^{1/3}-1 \sim 0.58$, as in Fig. 1.
This is the original DGP model on the tenth line of Table III,
which turns
out to be
indistinguishable from the flat DGP model on the ninth line.

About any isolated spherically symmetric
condensation of Schwarzschild radius $r_S:=2 G_N M/c^2$, the
self-accelerating metric \be g_{tt}=1-r_S/2r+\sqrt{r_S^2 r/2
r_{\star}^3},\qquad g_{rr}^{-1}=1+r_S/2 r-\sqrt{r_S^2 r/8
r_{\star}^3},\qquad r\lesssim r_{\star}, \ee so that Einstein
gravity obtains only up to the Vainstein scale \cite{DGZ} \be r_{\star}:=
(r_S r_c^2)^{1/3}\sim (H_0 r_S)^{1/3} H_0^{-1}\ll H_0^{-1}. \ee This
intermediate scale, $r_S\ll r_{\star} \ll H_0^{-1}$, is also where the
growth of fluctuations would change from Einstein gravity over to
linearized DGP or to scalar-tensor Brans-Dicke gravity, with an
effective Newton's constant slowly decreasing by no more than a
factor two \cite{LueScoccimaro}.  For cosmological scale $r_c \rightarrow \infty$, the
DGP modified Friedmann equation reduces to the Einstein-Friedmann
equation, but the DGP metric (C3) still does not reduce to the
Schwarzschild metric: There are no Iron Sphere or Birkhoff's
Theorems in DGP geometry.

The original flat DGP model can be generalized \cite{DvaliTurn} to
\be H^2-H^{\alpha} r_c^{\alpha-2}=\varkappa^2\rho/3, \qquad
1=\Omega_{m0}+\beta^{\alpha-2},\ee which is equivalent to a `dark
energy" $\rho_{DE}:= 3 M_P^2 H^2-\rho=3 M_P^2 H^{\alpha}
r_c^{\alpha-2},~w_{DE}=-1+\alpha/2$. This generalization reduces to
the original flat DGP form for $\alpha=1$, but otherwise
interpolates between the Einstein-de Sitter model for
$\alpha=2,~\beta=\infty $ and the Flat Classical Cosmological
Constant Model for $\alpha=0$. For small $\alpha$, it describes a
slowly varying cosmological constant.
\begin{acknowledgments}
I thank Richard Woodard (University of Florida) for helpful
discussions of $f(R)$ theories, and Dallas Kennedy
(MathWorks), Roy Maartens (Portsmouth) and Damien Easson (Durham)
for critical comments.
\end{acknowledgments}
\bibliography{bibliographyPhysrev} 

\begin{thebibliography}{81}
\expandafter\ifx\csname natexlab\endcsname\relax\def\natexlab#1{#1}\fi
\expandafter\ifx\csname bibnamefont\endcsname\relax
  \def\bibnamefont#1{#1}\fi
\expandafter\ifx\csname bibfnamefont\endcsname\relax
  \def\bibfnamefont#1{#1}\fi
\expandafter\ifx\csname citenamefont\endcsname\relax
  \def\citenamefont#1{#1}\fi
\expandafter\ifx\csname url\endcsname\relax
  \def\url#1{\texttt{#1}}\fi
\expandafter\ifx\csname urlprefix\endcsname\relax\def\urlprefix{URL }\fi
\providecommand{\bibinfo}[2]{#2}
\providecommand{\eprint}[2][]{\url{#2}}

\bibitem[{\citenamefont{Gu and Hwang}(2001)}]{Gu}
\bibinfo{author}{\bibfnamefont{J.-A.} \bibnamefont{Gu}} \bibnamefont{and}
  \bibinfo{author}{\bibfnamefont{W.-Y.} \bibnamefont{Hwang}},
  \bibinfo{journal}{Phys. Rev. D} \textbf{\bibinfo{volume}{65}},
  \bibinfo{pages}{024003} (\bibinfo{year}{2001}).

\bibitem[{\citenamefont{Sahni and Starobinsky}(2006)}]{Starobinsky}
\bibinfo{author}{\bibfnamefont{V.}~\bibnamefont{Sahni}} \bibnamefont{and}
  \bibinfo{author}{\bibfnamefont{A.~A.} \bibnamefont{Starobinsky}},
  \bibinfo{journal}{Int. J. Mod. Phys.} \textbf{\bibinfo{volume}{D15}},
  \bibinfo{pages}{2105} (\bibinfo{year}{2006}).

\bibitem[{\citenamefont{Davis et~al.}(2007)}]{Davis}
\bibinfo{author}{\bibfnamefont{T.~M.} \bibnamefont{Davis}} \bibnamefont{et~al.}
  (\bibinfo{year}{2007}), \bibinfo{note}{arXiv:astro-ph/0701510v1, Table 2,
  Fig. 7}.

\bibitem[{\citenamefont{Wood-Vasey et~al.}(2007)}]{Essence}
\bibinfo{author}{\bibfnamefont{W.}~\bibnamefont{Wood-Vasey}}
  \bibnamefont{et~al.}, \bibinfo{journal}{Astrophys. J}
  (\bibinfo{year}{2007}), \bibinfo{note}{[ESSENCE Program],
  arXiv:astro-ph/0701043v1}.

\bibitem[{\citenamefont{Dvali et~al.}(2000)\citenamefont{Dvali, Gabadadze, and
  Porrati}}]{DGP}
\bibinfo{author}{\bibfnamefont{G.}~\bibnamefont{Dvali}},
  \bibinfo{author}{\bibfnamefont{G.}~\bibnamefont{Gabadadze}},
  \bibnamefont{and} \bibinfo{author}{\bibfnamefont{M.}~\bibnamefont{Porrati}},
  \bibinfo{journal}{Phys. Lett. B} \textbf{\bibinfo{volume}{484}},
  \bibinfo{pages}{112} (\bibinfo{year}{2000}), \bibinfo{note}{[DGP]}.

\bibitem[{\citenamefont{Lue}(2006)}]{Lue}
\bibinfo{author}{\bibfnamefont{A.}~\bibnamefont{Lue}}, \bibinfo{journal}{Phys.
  Rep.} \textbf{\bibinfo{volume}{423}}, \bibinfo{pages}{1}
  (\bibinfo{year}{2006}).

\bibitem[{\citenamefont{Lue and Starkman}(2003)}]{LueStarkman}
\bibinfo{author}{\bibfnamefont{A.}~\bibnamefont{Lue}} \bibnamefont{and}
  \bibinfo{author}{\bibfnamefont{G.}~\bibnamefont{Starkman}},
  \bibinfo{journal}{Phys. Rev. D} \textbf{\bibinfo{volume}{67}},
  \bibinfo{pages}{064002} (\bibinfo{year}{2003}),
  \bibinfo{note}{arXiv:astro-ph/0212083}.

\bibitem[{\citenamefont{Iorio}(2005{\natexlab{a}})}]{Iorio}
\bibinfo{author}{\bibfnamefont{L.}~\bibnamefont{Iorio}}
  (\bibinfo{year}{2005}{\natexlab{a}}), \bibinfo{note}{arXiv:gr-qc/0510059}.

\bibitem[{\citenamefont{Iorio}(2005{\natexlab{b}})}]{IorioSecular}
\bibinfo{author}{\bibfnamefont{L.}~\bibnamefont{Iorio}}, \bibinfo{journal}{J.
  of Cosmology and Astrop. Physics} \textbf{\bibinfo{volume}{9}},
  \bibinfo{pages}{6} (\bibinfo{year}{2005}{\natexlab{b}}).

\bibitem[{\citenamefont{Ishak et~al.}(2006)\citenamefont{Ishak, Upadhye, and
  Spergel}}]{Ishak}
\bibinfo{author}{\bibfnamefont{M.}~\bibnamefont{Ishak}},
  \bibinfo{author}{\bibfnamefont{A.}~\bibnamefont{Upadhye}}, \bibnamefont{and}
  \bibinfo{author}{\bibfnamefont{D.}~\bibnamefont{Spergel}},
  \bibinfo{journal}{Phys. Rev. D} \textbf{\bibinfo{volume}{74}},
  \bibinfo{pages}{043513} (\bibinfo{year}{2006}),
  \bibinfo{note}{arXiv:astro-ph/0507184}.

\bibitem[{\citenamefont{Riess et~al.}(2007)}]{Riess2}
\bibinfo{author}{\bibfnamefont{A.}~\bibnamefont{Riess}} \bibnamefont{et~al.},
  \bibinfo{journal}{Astrophys. J} \textbf{\bibinfo{volume}{659}},
  \bibinfo{pages}{98} (\bibinfo{year}{2007}).

\bibitem[{\citenamefont{Spergel et~al.}(2006)}]{Spergel}
\bibinfo{author}{\bibfnamefont{D.}~\bibnamefont{Spergel}} \bibnamefont{et~al.}
  (\bibinfo{year}{2006}), \bibinfo{note}{arXiv:astro-ph/0603449, Tables 2, 9,
  [WMAP3]}.

\bibitem[{\citenamefont{Rozo et~al.}(2007)}]{clusters}
\bibinfo{author}{\bibfnamefont{E.}~\bibnamefont{Rozo}} \bibnamefont{et~al.},
  \bibinfo{journal}{Ap. J.}  (\bibinfo{year}{2007}),
  \bibinfo{note}{arXiv:astro-ph/0703571}.

\bibitem[{\citenamefont{Koyama and Maartens}(2006)}]{Koyama}
\bibinfo{author}{\bibfnamefont{K.}~\bibnamefont{Koyama}} \bibnamefont{and}
  \bibinfo{author}{\bibfnamefont{R.}~\bibnamefont{Maartens}},
  \bibinfo{journal}{JCAP} \textbf{\bibinfo{volume}{0601}}, \bibinfo{pages}{016}
  (\bibinfo{year}{2006}), \bibinfo{note}{arXiv:astro-ph/0511634}.

\bibitem[{\citenamefont{Lamoreaux}(2007)}]{Lamoreaux}
\bibinfo{author}{\bibfnamefont{S.~K.} \bibnamefont{Lamoreaux}},
  \bibinfo{journal}{Physics Today} \textbf{\bibinfo{volume}{60}},
  \bibinfo{pages}{40} (\bibinfo{year}{2007}).

\bibitem[{\citenamefont{Arkani-Hamad et~al.}(1998)\citenamefont{Arkani-Hamad,
  Dimopoulos, and Dvali}}]{Arkani-Hamed}
\bibinfo{author}{\bibfnamefont{N.}~\bibnamefont{Arkani-Hamad}},
  \bibinfo{author}{\bibfnamefont{S.}~\bibnamefont{Dimopoulos}},
  \bibnamefont{and} \bibinfo{author}{\bibfnamefont{G.}~\bibnamefont{Dvali}},
  \bibinfo{journal}{Phys. Lett. B} \textbf{\bibinfo{volume}{429}},
  \bibinfo{pages}{263} (\bibinfo{year}{1998}).

\bibitem[{\citenamefont{Randall and Sundrum}(1999)}]{Randall}
\bibinfo{author}{\bibfnamefont{L.}~\bibnamefont{Randall}} \bibnamefont{and}
  \bibinfo{author}{\bibfnamefont{R.}~\bibnamefont{Sundrum}},
  \bibinfo{journal}{Phys. Lett B} \textbf{\bibinfo{volume}{83}},
  \bibinfo{pages}{3370} (\bibinfo{year}{1999}).

\bibitem[{\citenamefont{Binutray et~al.}(2000)\citenamefont{Binutray, Deffayet,
  and Langlois}}]{Binutray}
\bibinfo{author}{\bibfnamefont{P.}~\bibnamefont{Binutray}},
  \bibinfo{author}{\bibfnamefont{C.}~\bibnamefont{Deffayet}}, \bibnamefont{and}
  \bibinfo{author}{\bibfnamefont{D.}~\bibnamefont{Langlois}},
  \bibinfo{journal}{Nucl. Phys. B} \textbf{\bibinfo{volume}{565}},
  \bibinfo{pages}{269} (\bibinfo{year}{2000}).

\bibitem[{\citenamefont{Lue et~al.}(2004)\citenamefont{Lue, Scoccimaro, and
  Starkman}}]{LueScoccimaro}
\bibinfo{author}{\bibfnamefont{A.}~\bibnamefont{Lue}},
  \bibinfo{author}{\bibfnamefont{R.}~\bibnamefont{Scoccimaro}},
  \bibnamefont{and} \bibinfo{author}{\bibfnamefont{G.}~\bibnamefont{Starkman}},
  \bibinfo{journal}{Phys. Rev. D} \textbf{\bibinfo{volume}{69}},
  \bibinfo{pages}{124015} (\bibinfo{year}{2004}).

\bibitem[{\citenamefont{Dvali et~al.}(2003)\citenamefont{Dvali, Gruzinov, and
  Zaldarriaga}}]{DGZ}
\bibinfo{author}{\bibfnamefont{G.}~\bibnamefont{Dvali}},
  \bibinfo{author}{\bibfnamefont{A.}~\bibnamefont{Gruzinov}}, \bibnamefont{and}
  \bibinfo{author}{\bibfnamefont{M.}~\bibnamefont{Zaldarriaga}},
  \bibinfo{journal}{Phys. Rev. D} \textbf{\bibinfo{volume}{68}},
  \bibinfo{pages}{024012} (\bibinfo{year}{2003}).

\bibitem[{\citenamefont{Sereno and Jetzer}(2006{\natexlab{a}})}]{Sereno}
\bibinfo{author}{\bibfnamefont{M.}~\bibnamefont{Sereno}} \bibnamefont{and}
  \bibinfo{author}{\bibfnamefont{P.}~\bibnamefont{Jetzer}},
  \bibinfo{journal}{MNRAS} \textbf{\bibinfo{volume}{371}}, \bibinfo{pages}{626}
  (\bibinfo{year}{2006}{\natexlab{a}}), \bibinfo{note}{arXiv:astro-ph/0606197}.

\bibitem[{\citenamefont{Milne}(1934)}]{Milne}
\bibinfo{author}{\bibfnamefont{E.}~\bibnamefont{Milne}},
  \bibinfo{journal}{Quart. J. Math. (Oxford)} \textbf{\bibinfo{volume}{5}},
  \bibinfo{pages}{64} (\bibinfo{year}{1934}).

\bibitem[{\citenamefont{McCrea and Milne}(1934)}]{McCrae}
\bibinfo{author}{\bibfnamefont{W.}~\bibnamefont{McCrea}} \bibnamefont{and}
  \bibinfo{author}{\bibfnamefont{E.}~\bibnamefont{Milne}},
  \bibinfo{journal}{Quart. J. Math. (Oxford)} \textbf{\bibinfo{volume}{5}},
  \bibinfo{pages}{73} (\bibinfo{year}{1934}).

\bibitem[{\citenamefont{Wang and Mukherjee}(2006)}]{WangMukherjee}
\bibinfo{author}{\bibfnamefont{Y.}~\bibnamefont{Wang}} \bibnamefont{and}
  \bibinfo{author}{\bibfnamefont{P.}~\bibnamefont{Mukherjee}},
  \bibinfo{journal}{Astrophys. J} \textbf{\bibinfo{volume}{650}},
  \bibinfo{pages}{1} (\bibinfo{year}{2006}).

\bibitem[{\citenamefont{Fairbairn and Goobar}(2006)}]{Fairbairn}
\bibinfo{author}{\bibfnamefont{M.}~\bibnamefont{Fairbairn}} \bibnamefont{and}
  \bibinfo{author}{\bibfnamefont{A.}~\bibnamefont{Goobar}},
  \bibinfo{journal}{Physics Letters B} \textbf{\bibinfo{volume}{642}},
  \bibinfo{pages}{432} (\bibinfo{year}{2006}).

\bibitem[{\citenamefont{Eisenstein et~al.}(2005)}]{Eisenstein}
\bibinfo{author}{\bibfnamefont{D.}~\bibnamefont{Eisenstein}}
  \bibnamefont{et~al.}, \bibinfo{journal}{Astrophys. J.}
  \textbf{\bibinfo{volume}{633}}, \bibinfo{pages}{560} (\bibinfo{year}{2005}),
  \bibinfo{note}{[SDSS Luminous Red Galaxy Survey]}.

\bibitem[{\citenamefont{Perlmutter et~al.}(1999)}]{SCP}
\bibinfo{author}{\bibfnamefont{S.}~\bibnamefont{Perlmutter}}
  \bibnamefont{et~al.}, \bibinfo{journal}{Astrophys. J.}
  \textbf{\bibinfo{volume}{517}}, \bibinfo{pages}{565} (\bibinfo{year}{1999}),
  \bibinfo{note}{[Supernova Cosmology Project]}.

\bibitem[{\citenamefont{Astier et~al.}(2006)}]{SNLS}
\bibinfo{author}{\bibfnamefont{P.}~\bibnamefont{Astier}} \bibnamefont{et~al.},
  \bibinfo{journal}{Astron. and Astroph.} \textbf{\bibinfo{volume}{447}},
  \bibinfo{pages}{31} (\bibinfo{year}{2006}), \bibinfo{note}{[Supernova Legacy
  Survey SNLS]}.

\bibitem[{\citenamefont{Miknaitis et~al.}(2007)}]{Miknaitis}
\bibinfo{author}{\bibfnamefont{G.}~\bibnamefont{Miknaitis}}
  \bibnamefont{et~al.}, \bibinfo{journal}{Ap. J.}  (\bibinfo{year}{2007}),
  \bibinfo{note}{arXiv:astro-ph/0701043}.

\bibitem[{\citenamefont{Wang and Tegmark}(2004)}]{Tegmark3}
\bibinfo{author}{\bibfnamefont{Y.}~\bibnamefont{Wang}} \bibnamefont{and}
  \bibinfo{author}{\bibfnamefont{M.}~\bibnamefont{Tegmark}},
  \bibinfo{journal}{Phys. Rev. Lett.} \textbf{\bibinfo{volume}{92}},
  \bibinfo{pages}{241302} (\bibinfo{year}{2004}).

\bibitem[{\citenamefont{Maor et~al.}(2001)\citenamefont{Maor, Brustein, and
  Steinhardt}}]{Moar}
\bibinfo{author}{\bibfnamefont{L.}~\bibnamefont{Maor}},
  \bibinfo{author}{\bibfnamefont{R.}~\bibnamefont{Brustein}}, \bibnamefont{and}
  \bibinfo{author}{\bibfnamefont{P.}~\bibnamefont{Steinhardt}},
  \bibinfo{journal}{Phys. Rev. Lett.}  (\bibinfo{year}{2001}).

\bibitem[{\citenamefont{Linder and Huterer}(2005)}]{LindHut}
\bibinfo{author}{\bibfnamefont{E.}~\bibnamefont{Linder}} \bibnamefont{and}
  \bibinfo{author}{\bibfnamefont{D.}~\bibnamefont{Huterer}},
  \bibinfo{journal}{Phys. Rev. D} \textbf{\bibinfo{volume}{72}},
  \bibinfo{pages}{043509} (\bibinfo{year}{2005}).

\bibitem[{\citenamefont{Caldwell and Linder}(2005)}]{CaldLinder}
\bibinfo{author}{\bibfnamefont{R.~R.} \bibnamefont{Caldwell}} \bibnamefont{and}
  \bibinfo{author}{\bibfnamefont{E.~V.} \bibnamefont{Linder}},
  \bibinfo{journal}{Phys. Rev. Lett.} \textbf{\bibinfo{volume}{95}},
  \bibinfo{pages}{141301} (\bibinfo{year}{2005}).

\bibitem[{\citenamefont{Chevalier and Polarski}(2001)}]{Polarski2001}
\bibinfo{author}{\bibfnamefont{M.}~\bibnamefont{Chevalier}} \bibnamefont{and}
  \bibinfo{author}{\bibfnamefont{D.}~\bibnamefont{Polarski}},
  \bibinfo{journal}{Int. J. Mod. Phys.} \textbf{\bibinfo{volume}{D10}},
  \bibinfo{pages}{213} (\bibinfo{year}{2001}), \bibinfo{note}{arXiv
  gr-qc/0009008}.

\bibitem[{\citenamefont{Linder}(2003)}]{Linder}
\bibinfo{author}{\bibfnamefont{E.}~\bibnamefont{Linder}},
  \bibinfo{journal}{Phys. Rev. Lett.} \textbf{\bibinfo{volume}{90}},
  \bibinfo{pages}{091301} (\bibinfo{year}{2003}).

\bibitem[{\citenamefont{Wang and Freese}(2006)}]{WangFreese}
\bibinfo{author}{\bibfnamefont{Y.}~\bibnamefont{Wang}} \bibnamefont{and}
  \bibinfo{author}{\bibfnamefont{K.}~\bibnamefont{Freese}},
  \bibinfo{journal}{Phys. Lett. B} \textbf{\bibinfo{volume}{632}},
  \bibinfo{pages}{449} (\bibinfo{year}{2006}).

\bibitem[{\citenamefont{Buchmuller and Dragon}(1988)}]{Buch}
\bibinfo{author}{\bibfnamefont{W.}~\bibnamefont{Buchmuller}} \bibnamefont{and}
  \bibinfo{author}{\bibfnamefont{N.}~\bibnamefont{Dragon}},
  \bibinfo{journal}{Phys. Lett. B} \textbf{\bibinfo{volume}{207}},
  \bibinfo{pages}{292} (\bibinfo{year}{1988}).

\bibitem[{\citenamefont{Unruh}(1989)}]{Unruh}
\bibinfo{author}{\bibfnamefont{W.}~\bibnamefont{Unruh}},
  \bibinfo{journal}{Phys. Rev. D} \textbf{\bibinfo{volume}{40}},
  \bibinfo{pages}{1048} (\bibinfo{year}{1989}).

\bibitem[{\citenamefont{Melchiorri et~al.}(2007)\citenamefont{Melchiorri,
  Pagano, and Pandolfi}}]{AccelStart}
\bibinfo{author}{\bibfnamefont{A.}~\bibnamefont{Melchiorri}},
  \bibinfo{author}{\bibfnamefont{L.}~\bibnamefont{Pagano}}, \bibnamefont{and}
  \bibinfo{author}{\bibfnamefont{S.}~\bibnamefont{Pandolfi}}
  (\bibinfo{year}{2007}), \bibinfo{note}{arXiv:0706.1314 [astro-ph]}.

\bibitem[{\citenamefont{Gondolo and Freese}(2003)}]{Freese}
\bibinfo{author}{\bibfnamefont{P.}~\bibnamefont{Gondolo}} \bibnamefont{and}
  \bibinfo{author}{\bibfnamefont{K.}~\bibnamefont{Freese}},
  \bibinfo{journal}{Phys. Rev. D} \textbf{\bibinfo{volume}{68}},
  \bibinfo{pages}{063509} (\bibinfo{year}{2003}).

\bibitem[{\citenamefont{Szydlowski et~al.}(2006)\citenamefont{Szydlowski,
  Kurek, and Krawiec}}]{DEfits}
\bibinfo{author}{\bibfnamefont{M.}~\bibnamefont{Szydlowski}},
  \bibinfo{author}{\bibfnamefont{A.}~\bibnamefont{Kurek}}, \bibnamefont{and}
  \bibinfo{author}{\bibfnamefont{A.}~\bibnamefont{Krawiec}},
  \bibinfo{journal}{Phys. Lett. B} \textbf{\bibinfo{volume}{642}},
  \bibinfo{pages}{206} (\bibinfo{year}{2006}),
  \bibinfo{note}{arXiv:astro-ph/0604327 v2}.

\bibitem[{\citenamefont{Albrecht et~al.}(2006)}]{DETF}
\bibinfo{author}{\bibfnamefont{A.}~\bibnamefont{Albrecht}} \bibnamefont{et~al.}
  (\bibinfo{year}{2006}), \bibinfo{note}{arXiv:astro-ph/0609591, [DETF]}.

\bibitem[{\citenamefont{Padmanabhan}(2005)}]{Padmanabhan}
\bibinfo{author}{\bibfnamefont{T.}~\bibnamefont{Padmanabhan}},
  \bibinfo{journal}{Curr. Sci.} \textbf{\bibinfo{volume}{88}},
  \bibinfo{pages}{1057} (\bibinfo{year}{2005}).

\bibitem[{\citenamefont{Bonvin et~al.}(2006)\citenamefont{Bonvin, Caprini, and
  Durrer}}]{Durrer}
\bibinfo{author}{\bibfnamefont{C.}~\bibnamefont{Bonvin}},
  \bibinfo{author}{\bibfnamefont{C.}~\bibnamefont{Caprini}}, \bibnamefont{and}
  \bibinfo{author}{\bibfnamefont{R.}~\bibnamefont{Durrer}},
  \bibinfo{journal}{Phys. Rev. Lett.} \textbf{\bibinfo{volume}{97}},
  \bibinfo{pages}{081303} (\bibinfo{year}{2006}),
  \bibinfo{note}{arXiv:astro-ph/0606584}.

\bibitem[{\citenamefont{Bento et~al.}(2005)\citenamefont{Bento, Bertolami,
  Santos, and Sen}}]{Bento}
\bibinfo{author}{\bibfnamefont{M.~C.} \bibnamefont{Bento}},
  \bibinfo{author}{\bibfnamefont{O.}~\bibnamefont{Bertolami}},
  \bibinfo{author}{\bibfnamefont{N.~M.~C.} \bibnamefont{Santos}},
  \bibnamefont{and} \bibinfo{author}{\bibfnamefont{A.}~\bibnamefont{Sen}},
  \bibinfo{journal}{Phys. Rev. D}  (\bibinfo{year}{2005}),
  \bibinfo{note}{arXiv:astro-ph/0412638}.

\bibitem[{\citenamefont{Amendola et~al.}(2003)\citenamefont{Amendola, Finelli,
  Burigana, and Carturan}}]{Amendola2003}
\bibinfo{author}{\bibfnamefont{L.}~\bibnamefont{Amendola}},
  \bibinfo{author}{\bibfnamefont{F.}~\bibnamefont{Finelli}},
  \bibinfo{author}{\bibfnamefont{C.}~\bibnamefont{Burigana}}, \bibnamefont{and}
  \bibinfo{author}{\bibfnamefont{D.}~\bibnamefont{Carturan}},
  \bibinfo{journal}{JCAP} \textbf{\bibinfo{volume}{309}}, \bibinfo{pages}{5}
  (\bibinfo{year}{2003}), \bibinfo{note}{arXiv:astro-ph/0304325}.

\bibitem[{\citenamefont{Zhu}(2004)}]{Zhu}
\bibinfo{author}{\bibfnamefont{Z.-H.} \bibnamefont{Zhu}},
  \bibinfo{journal}{Astronomy and Astrophysics}  (\bibinfo{year}{2004}).

\bibitem[{\citenamefont{Deffayet}(2001)}]{Deffayet2001}
\bibinfo{author}{\bibfnamefont{C.}~\bibnamefont{Deffayet}},
  \bibinfo{journal}{Phys. Lett. B} \textbf{\bibinfo{volume}{502}},
  \bibinfo{pages}{199} (\bibinfo{year}{2001}),
  \bibinfo{note}{arXiv:hep-ph/0010186}.

\bibitem[{\citenamefont{Deffayet et~al.}(2002)}]{Deffayet2}
\bibinfo{author}{\bibfnamefont{C.}~\bibnamefont{Deffayet}}
  \bibnamefont{et~al.}, \bibinfo{journal}{Phys. Rev. D}
  \textbf{\bibinfo{volume}{66}}, \bibinfo{pages}{024019}
  (\bibinfo{year}{2002}).

\bibitem[{\citenamefont{Hockstra et~al.}(2006)}]{CFHT}
\bibinfo{author}{\bibfnamefont{H.}~\bibnamefont{Hockstra}}
  \bibnamefont{et~al.}, \bibinfo{journal}{Astrophys. J.}
  \textbf{\bibinfo{volume}{647}}, \bibinfo{pages}{116} (\bibinfo{year}{2006}),
  \bibinfo{note}{[Canada-France-Hawaii Telescope]}.

\bibitem[{\citenamefont{Ellis}(2006)}]{Ellis}
\bibinfo{author}{\bibfnamefont{G.}~\bibnamefont{Ellis}},
  \emph{\bibinfo{title}{Handbook of Philosophy of Physics}}
  (\bibinfo{publisher}{Elsevier}, \bibinfo{year}{2006}).

\bibitem[{\citenamefont{Bludman and Ross}(2002)}]{BludRoss}
\bibinfo{author}{\bibfnamefont{S.}~\bibnamefont{Bludman}} \bibnamefont{and}
  \bibinfo{author}{\bibfnamefont{M.}~\bibnamefont{Ross}},
  \bibinfo{journal}{Phys. Rev. D} \textbf{\bibinfo{volume}{65}},
  \bibinfo{pages}{043503} (\bibinfo{year}{2002}).

\bibitem[{\citenamefont{Smolin}(2006)}]{Smolin}
\bibinfo{author}{\bibfnamefont{L.}~\bibnamefont{Smolin}},
  \emph{\bibinfo{title}{The Trouble with Physics}}
  (\bibinfo{publisher}{Houghton Mifflin}, \bibinfo{year}{2006}), ISBN
  \bibinfo{isbn}{13: 978-0-618-55105-7}.

\bibitem[{\citenamefont{Weinberg}(2006)}]{WeinbergAnthro}
\bibinfo{author}{\bibfnamefont{S.}~\bibnamefont{Weinberg}},
  \emph{\bibinfo{title}{Universe or Multiverse?,ed. B. Carr}}
  (\bibinfo{publisher}{Cambridge University Press}, \bibinfo{year}{2006}),
  \bibinfo{note}{arXiv:hep-th/0511037}.

\bibitem[{\citenamefont{Damour}(2004)}]{Damour}
\bibinfo{author}{\bibfnamefont{T.}~\bibnamefont{Damour}},
  \bibinfo{journal}{Phys. Lett. B} \textbf{\bibinfo{volume}{592}},
  \bibinfo{pages}{186} (\bibinfo{year}{2004}), \bibinfo{note}{[Review of
  Particle Physics]}.

\bibitem[{\citenamefont{Gabadadze and Iglesias}(2006)}]{Gabadadze}
\bibinfo{author}{\bibfnamefont{G.}~\bibnamefont{Gabadadze}} \bibnamefont{and}
  \bibinfo{author}{\bibfnamefont{A.}~\bibnamefont{Iglesias}},
  \bibinfo{journal}{Phys. Lett. B} \textbf{\bibinfo{volume}{632}},
  \bibinfo{pages}{617} (\bibinfo{year}{2006}).

\bibitem[{\citenamefont{Sereno and Jetzer}(2006{\natexlab{b}})}]{Sereno2006}
\bibinfo{author}{\bibfnamefont{M.}~\bibnamefont{Sereno}} \bibnamefont{and}
  \bibinfo{author}{\bibfnamefont{P.}~\bibnamefont{Jetzer}},
  \bibinfo{journal}{Phys. Rev. D} \textbf{\bibinfo{volume}{73}},
  \bibinfo{pages}{063004} (\bibinfo{year}{2006}{\natexlab{b}}),
  \bibinfo{note}{arXiv:astro-ph/0602438}.

\bibitem[{\citenamefont{Damour and
  Polyakov}(1994{\natexlab{a}})}]{DamourPolyakov}
\bibinfo{author}{\bibfnamefont{T.}~\bibnamefont{Damour}} \bibnamefont{and}
  \bibinfo{author}{\bibfnamefont{A.}~\bibnamefont{Polyakov}},
  \bibinfo{journal}{Nucl. Phys. B}  (\bibinfo{year}{1994}{\natexlab{a}}).

\bibitem[{\citenamefont{Damour and
  Polyakov}(1994{\natexlab{b}})}]{DamourPolyakov2}
\bibinfo{author}{\bibfnamefont{T.}~\bibnamefont{Damour}} \bibnamefont{and}
  \bibinfo{author}{\bibfnamefont{A.}~\bibnamefont{Polyakov}},
  \bibinfo{journal}{Gen. Relativ. Gravit.}
  (\bibinfo{year}{1994}{\natexlab{b}}).

\bibitem[{\citenamefont{Fujii and Maeda}(2003)}]{Fujii}
\bibinfo{author}{\bibfnamefont{Y.}~\bibnamefont{Fujii}} \bibnamefont{and}
  \bibinfo{author}{\bibfnamefont{K.-I.} \bibnamefont{Maeda}},
  \emph{\bibinfo{title}{The Scalar-Tensor Theory of Gravitation}}
  (\bibinfo{publisher}{Cambridge University Press, Cambridge, England},
  \bibinfo{year}{2003}).

\bibitem[{\citenamefont{Capozziello et~al.}(2005)\citenamefont{Capozziello,
  Cardone, and Troisi}}]{Cap2005}
\bibinfo{author}{\bibfnamefont{S.}~\bibnamefont{Capozziello}},
  \bibinfo{author}{\bibfnamefont{V.~F.} \bibnamefont{Cardone}},
  \bibnamefont{and} \bibinfo{author}{\bibfnamefont{A.}~\bibnamefont{Troisi}},
  \bibinfo{journal}{Phys. Rev. D} \textbf{\bibinfo{volume}{71}},
  \bibinfo{pages}{043503} (\bibinfo{year}{2005}).

\bibitem[{\citenamefont{Boisseau et~al.}(2000)\citenamefont{Boisseau,
  Esposito-Farese, Polarski, and Starobinsky}}]{Starobinsky2000}
\bibinfo{author}{\bibfnamefont{B.}~\bibnamefont{Boisseau}},
  \bibinfo{author}{\bibfnamefont{G.}~\bibnamefont{Esposito-Farese}},
  \bibinfo{author}{\bibfnamefont{D.}~\bibnamefont{Polarski}}, \bibnamefont{and}
  \bibinfo{author}{\bibfnamefont{A.~A.} \bibnamefont{Starobinsky}},
  \bibinfo{journal}{Phys. Rev. Lett.} \textbf{\bibinfo{volume}{85}},
  \bibinfo{pages}{2236} (\bibinfo{year}{2000}),
  \bibinfo{note}{arXiv:gr-qc/0001066}.

\bibitem[{\citenamefont{Bertotti et~al.}(2003)\citenamefont{Bertotti, Das, and
  Tortora}}]{Bertotti}
\bibinfo{author}{\bibfnamefont{B.}~\bibnamefont{Bertotti}},
  \bibinfo{author}{\bibfnamefont{L.}~\bibnamefont{Das}}, \bibnamefont{and}
  \bibinfo{author}{\bibfnamefont{P.}~\bibnamefont{Tortora}},
  \bibinfo{journal}{Nature} \textbf{\bibinfo{volume}{425}},
  \bibinfo{pages}{374} (\bibinfo{year}{2003}).

\bibitem[{\citenamefont{Capozziello
  et~al.}(2006{\natexlab{a}})\citenamefont{Capozziello, Cardone, and
  Troisi}}]{Cap2006}
\bibinfo{author}{\bibfnamefont{S.}~\bibnamefont{Capozziello}},
  \bibinfo{author}{\bibfnamefont{V.~P.} \bibnamefont{Cardone}},
  \bibnamefont{and} \bibinfo{author}{\bibfnamefont{A.}~\bibnamefont{Troisi}},
  \bibinfo{journal}{JCAP} \textbf{\bibinfo{volume}{0608}}, \bibinfo{pages}{001}
  (\bibinfo{year}{2006}{\natexlab{a}}), \bibinfo{note}{arXiv:astro-ph/0602349}.

\bibitem[{\citenamefont{Catena et~al.}(2004)\citenamefont{Catena, Fornengo,
  Masiero, Pietroni, and Rosati}}]{Catena}
\bibinfo{author}{\bibfnamefont{R.}~\bibnamefont{Catena}},
  \bibinfo{author}{\bibfnamefont{N.}~\bibnamefont{Fornengo}},
  \bibinfo{author}{\bibfnamefont{A.}~\bibnamefont{Masiero}},
  \bibinfo{author}{\bibfnamefont{M.}~\bibnamefont{Pietroni}}, \bibnamefont{and}
  \bibinfo{author}{\bibfnamefont{F.}~\bibnamefont{Rosati}}
  (\bibinfo{year}{2004}), \bibinfo{note}{arXiv:astro-ph/0406152}.

\bibitem[{\citenamefont{Ostrogradski}(1850)}]{Ostrogradski}
\bibinfo{author}{\bibfnamefont{M.}~\bibnamefont{Ostrogradski}},
  \bibinfo{journal}{Mem. Ac. St. Petersbourg} \textbf{\bibinfo{volume}{VI 4}},
  \bibinfo{pages}{385} (\bibinfo{year}{1850}).

\bibitem[{\citenamefont{Woodard}(2006)}]{Woodard2006}
\bibinfo{author}{\bibfnamefont{R.}~\bibnamefont{Woodard}}
  (\bibinfo{year}{2006}), \bibinfo{note}{arXiv:astro-ph/0601672}.

\bibitem[{\citenamefont{Carroll et~al.}(2004)\citenamefont{Carroll, Duvvuri,
  Trodden, and Turner}}]{Carroll}
\bibinfo{author}{\bibfnamefont{S.~M.} \bibnamefont{Carroll}},
  \bibinfo{author}{\bibfnamefont{V.}~\bibnamefont{Duvvuri}},
  \bibinfo{author}{\bibfnamefont{M.}~\bibnamefont{Trodden}}, \bibnamefont{and}
  \bibinfo{author}{\bibfnamefont{M.~S.} \bibnamefont{Turner}},
  \bibinfo{journal}{Phys. Rev. D} \textbf{\bibinfo{volume}{70}},
  \bibinfo{pages}{043528} (\bibinfo{year}{2004}).

\bibitem[{\citenamefont{Carroll et~al.}(2005)\citenamefont{Carroll, Felice,
  Duvvuri, Easson, Trodden, and Turner}}]{Carroll2}
\bibinfo{author}{\bibfnamefont{S.~M.} \bibnamefont{Carroll}},
  \bibinfo{author}{\bibfnamefont{A.~D.} \bibnamefont{Felice}},
  \bibinfo{author}{\bibfnamefont{V.}~\bibnamefont{Duvvuri}},
  \bibinfo{author}{\bibfnamefont{D.~A.} \bibnamefont{Easson}},
  \bibinfo{author}{\bibfnamefont{M.}~\bibnamefont{Trodden}}, \bibnamefont{and}
  \bibinfo{author}{\bibfnamefont{M.~S.} \bibnamefont{Turner}},
  \bibinfo{journal}{Phys. Rev. D} \textbf{\bibinfo{volume}{71}},
  \bibinfo{pages}{063513} (\bibinfo{year}{2005}),
  \bibinfo{note}{arXiv:astro-ph/0410031}.

\bibitem[{\citenamefont{Nojiri and Odintsov}(2007)}]{Nojiri}
\bibinfo{author}{\bibfnamefont{S.}~\bibnamefont{Nojiri}} \bibnamefont{and}
  \bibinfo{author}{\bibfnamefont{S.}~\bibnamefont{Odintsov}},
  \bibinfo{journal}{Int. J. Geom. Meth. Mod. Phys.}
  \textbf{\bibinfo{volume}{4}}, \bibinfo{pages}{115} (\bibinfo{year}{2007}),
  \bibinfo{note}{arXiv:hep-th/0601213}.

\bibitem[{\citenamefont{Soussa and Woodard}(2004)}]{Soussa}
\bibinfo{author}{\bibfnamefont{M.}~\bibnamefont{Soussa}} \bibnamefont{and}
  \bibinfo{author}{\bibfnamefont{R.}~\bibnamefont{Woodard}},
  \bibinfo{journal}{Gen. Relativ. Gravit.} \textbf{\bibinfo{volume}{36}},
  \bibinfo{pages}{855} (\bibinfo{year}{2004}).

\bibitem[{\citenamefont{Dolgov and Kawasaki}(2003)}]{Dolgov}
\bibinfo{author}{\bibfnamefont{A.~D.} \bibnamefont{Dolgov}} \bibnamefont{and}
  \bibinfo{author}{\bibfnamefont{M.}~\bibnamefont{Kawasaki}},
  \bibinfo{journal}{Phys. Lett. B} \textbf{\bibinfo{volume}{573}},
  \bibinfo{pages}{1} (\bibinfo{year}{2003}).

\bibitem[{\citenamefont{Teyssandier and Tourrenc}(1983)}]{Teyssandier}
\bibinfo{author}{\bibfnamefont{P.}~\bibnamefont{Teyssandier}} \bibnamefont{and}
  \bibinfo{author}{\bibfnamefont{P.}~\bibnamefont{Tourrenc}},
  \bibinfo{journal}{J. Math. Phys.}  (\bibinfo{year}{1983}).

\bibitem[{\citenamefont{Olmo}(2005)}]{Olmo}
\bibinfo{author}{\bibfnamefont{G.}~\bibnamefont{Olmo}}, \bibinfo{journal}{Phys.
  Rev. D} \textbf{\bibinfo{volume}{72}}, \bibinfo{pages}{083505}
  (\bibinfo{year}{2005}).

\bibitem[{\citenamefont{Capozziello
  et~al.}(2006{\natexlab{b}})\citenamefont{Capozziello, Nojiri, and
  Odintsov}}]{CapNojOdin}
\bibinfo{author}{\bibfnamefont{S.}~\bibnamefont{Capozziello}},
  \bibinfo{author}{\bibfnamefont{S.}~\bibnamefont{Nojiri}}, \bibnamefont{and}
  \bibinfo{author}{\bibfnamefont{S.~D.} \bibnamefont{Odintsov}},
  \bibinfo{journal}{Phys. Lett. B} \textbf{\bibinfo{volume}{634}},
  \bibinfo{pages}{93} (\bibinfo{year}{2006}{\natexlab{b}}),
  \bibinfo{note}{arXiv:hep-th/0512118}.

\bibitem[{\citenamefont{Nojiri and Odintsov}(2004)}]{Odintsov2}
\bibinfo{author}{\bibfnamefont{S.}~\bibnamefont{Nojiri}} \bibnamefont{and}
  \bibinfo{author}{\bibfnamefont{S.}~\bibnamefont{Odintsov}},
  \bibinfo{journal}{Gen. Relativ. Gravit.} \textbf{\bibinfo{volume}{36}},
  \bibinfo{pages}{1765} (\bibinfo{year}{2004}).

\bibitem[{\citenamefont{Nojiri and Odintsov}(2003)}]{Odintsov3}
\bibinfo{author}{\bibfnamefont{S.}~\bibnamefont{Nojiri}} \bibnamefont{and}
  \bibinfo{author}{\bibfnamefont{S.}~\bibnamefont{Odintsov}},
  \bibinfo{journal}{Phys. Rev. D} \textbf{\bibinfo{volume}{68}},
  \bibinfo{pages}{123512} (\bibinfo{year}{2003}).

\bibitem[{\citenamefont{Amendola et~al.}(2007)\citenamefont{Amendola, Polarski,
  and Tsujikawa}}]{Amendola2}
\bibinfo{author}{\bibfnamefont{L.}~\bibnamefont{Amendola}},
  \bibinfo{author}{\bibfnamefont{D.}~\bibnamefont{Polarski}}, \bibnamefont{and}
  \bibinfo{author}{\bibfnamefont{S.}~\bibnamefont{Tsujikawa}},
  \bibinfo{journal}{Phys. Rev. Lett.} \textbf{\bibinfo{volume}{98}},
  \bibinfo{pages}{131302} (\bibinfo{year}{2007}).

\bibitem[{\citenamefont{Bekenstein}(2004)}]{Bek}
\bibinfo{author}{\bibfnamefont{J.~D.} \bibnamefont{Bekenstein}},
  \bibinfo{journal}{Phys. Rev. D} \textbf{\bibinfo{volume}{70}},
  \bibinfo{pages}{083509} (\bibinfo{year}{2004}).

\bibitem[{\citenamefont{Skordis et~al.}(2006)\citenamefont{Skordis, Mota,
  Ferreira, and Boehm}}]{Skordis}
\bibinfo{author}{\bibfnamefont{C.}~\bibnamefont{Skordis}},
  \bibinfo{author}{\bibfnamefont{D.}~\bibnamefont{Mota}},
  \bibinfo{author}{\bibfnamefont{P.}~\bibnamefont{Ferreira}}, \bibnamefont{and}
  \bibinfo{author}{\bibfnamefont{C.}~\bibnamefont{Boehm}},
  \bibinfo{journal}{Phys.Rev.Lett.} \textbf{\bibinfo{volume}{96}},
  \bibinfo{pages}{011301} (\bibinfo{year}{2006}),
  \bibinfo{note}{arXiv:astro-ph/0505519}.

\bibitem[{\citenamefont{Dvali and Turner}(2003)}]{DvaliTurn}
\bibinfo{author}{\bibfnamefont{G.}~\bibnamefont{Dvali}} \bibnamefont{and}
  \bibinfo{author}{\bibfnamefont{M.}~\bibnamefont{Turner}}
  (\bibinfo{year}{2003}), \bibinfo{note}{arXiv:astro-ph/0301510}.

\end{thebibliography}
\end{document}